\documentclass[twoside,twocolumn]{article}

\usepackage{blindtext} 

\usepackage[sc]{mathpazo} 
\usepackage[T1]{fontenc} 
\usepackage{microtype} 

\usepackage[english]{babel} 

\usepackage[hmarginratio=1:1,top=32mm,columnsep=20pt]{geometry} 
\usepackage[hang, small,labelfont=bf,up,textfont=it,up]{caption} 
\usepackage{booktabs} 
\usepackage{color}
\usepackage{lettrine} 

\usepackage{enumitem} 
\setlist[itemize]{noitemsep} 

\usepackage{abstract} 

\usepackage{titlesec} 
\renewcommand\thesection{\Roman{section}} 
\renewcommand\thesubsection{\roman{subsection}} 
\titleformat{\section}[block]{\large\scshape\centering}{\thesection.}{1em}{} 
\titleformat{\subsection}[block]{\large}{\thesubsection.}{1em}{} 

\usepackage{fancyhdr} 
\usepackage{titling} 

\usepackage{graphicx}
\usepackage{caption}
\usepackage{subcaption}
\usepackage{siunitx}
\usepackage{hyperref} 

\pretitle{\begin{center}\Huge\bfseries} 
\posttitle{\end{center}} 
\title{A "DIY" data acquisition system for acoustic field measurements under harsh conditions} 
\author{%
\textsc{S. Büchholz, M. Lemke, J. Reiss and J.L. Sesterhenn} \\[1ex]
\normalsize Technische Universität Berlin \\ 
\normalsize \href{mailto:steffen.buechholz@tnt.tu-berlin.de}{steffen.buechholz@tnt.tu-berlin.de} 
}
\date{\today} 
\renewcommand{\maketitlehookd}{%
\begin{abstract}
\noindent 
Monitoring active volcanos is an ongoing and important task helping to understand and predict volcanic eruptions. In recent years, analysing the acoustic properties of eruptions became more relevant. We present an inexpensive, lightweight, portable, easy to use and modular acoustic data acquisition system for field measurements that can record data with up to 100~kHz. The system is based on a Raspberry Pi 3 B running a custom build bare metal operating system. It connects to an external analog - digital converter with the microphone sensor. A GPS receiver allows the logging of the position and in addition the recording of a very accurate time signal synchronously to the acoustic data. With that, it is possible for multiple modules to effectively work as a single microphone array. The whole system can be build with low cost and demands only minimal technical infrastructure. We demonstrate a possible use of such a microphone array by deploying 20 modules on the active volcano \textit{Stromboli} in the Aeolian Islands by Sicily, Italy. We use the collected acoustic data to indentify the sound source position for all recorded eruptions.
\end{abstract}
}

\begin{document}

\maketitle


\section{Introduction}
Geophysical data collection is an important aspect of monitoring volcanos. Infrasound is historically a very common quantity to record and analyse (\cite{Johnson:2011}, \cite{Matoza:2009}). Volcanic activity produces a lot of infrasound, so recording and analysing it gives useful insights in understanding and predicting (major) volcanic eruptions. However, analysing the acoustic data from volcanic eruptions that is in the audible sound spectrum and above (simply referred to as "acoustic data" from here on) recently came more into focus (\cite{Fernandez:2020}, \cite{Goto:2014}, \cite{Taddeucci:2014}). Since different kinds of eruptions also have very different acoustic characteristics, the acoustic data in this spectrum is also useful for analysing the behaviour of the volcano. It allows the eruptions to be categorized by their acoustic properties and can support to identify and track shifts in the eruptive behaviour. Another application, while not exclusive to the audible range, is the ability to localize the eruptions. For this, synchronous acoustic data from different measuring locations is needed.
In this paper, we demonstrate a custom data acquisition system for acoustic data. The system consists of modules that are based on a Raspberry Pi 3 B \cite{rpi_3B}. Connected to it is an analog - digital converter (ADC), a GPS receiver and an optional sensor for temperature and ambient pressure. The system is powered by a simple power bank, as often used to charge mobile phones. We build 20 prototype modules and deploy them on the active volcano \textit{Stromboli} in the Aeolian Islands by Sicily, Italy. A similar approach for a data acquisition system has been presented before by developing the GEM infrasound data logger \cite{Anderson:2018}, which exclusively logs infrasound. It is based on the Arduino system. Due to the much higher sampling frequency that is necessary to sample acoustic data in the audible spectrum and above, the data acquisition we develop needs to meet higher requirements in different aspects. Thus, we build a new system based on the Raspberry Pi platform, which offers more performance to meet these requirements.\\
The paper is organized as follows: In section \ref{sec:DAQ}, we explain the design of the system on a high level. Section \ref{sec:ExpSetup} gives some more details on the measurement campaign and in section \ref{sec:Results} we use the collected data to identify the location of all the recorded eruptions, more specifically the location of the sound sources using a simplified Time Difference of Arrival (TDoA) method \cite{TDoA}.


\section{Data acquisition system}\label{sec:DAQ}
We develop a prototype for a lightweight, easy to use in-field data acquisition system for acoustic measurements. The system consists of a module that records a single microphone signal together with the Pulse Per Second (PPS) signal from a GPS receiver. The high accuracy of the PPS signal allows multiple modules to effectively act as a single microphone array, simply by running them in parallel. Time consuming wiring and manual position measurements are avoided this way. Each module is capable of recording acoustic data with a sample rate of up to 100kHz. The synchronization between the modules is realized offline in post processing by synchronizing the acoustic signal via the recorded PPS signal. The PPS signal is received through an integrated GPS receiver. The main control unit of each module is a Raspberry Pi. Specifically the model Raspberry Pi 3 B (rpi3B), even though it should be mentioned that, if needed, porting the system to other models of the raspberry pi family can be done easily. The multicore architecture of the rpi3B allows to separate and distribute tasks between cores. Especially the collection of the acoustic data samples at an accurate high sample rate requires special care and is dedicated exclusively to one of the 4 cores of the rpi3B. Since the linux based operating system (OS) that would usually run on the rpi3B does not allow the low level control over threads and timings, a custom bare metal real time OS was developed using the tool ultibo \cite{ultibo}, a bare metal development environment. Since the rpi3B does not contain any analog-digital converter (ADC) to record the signals from a microphone sensor, an external ADC is used that interfaces with the rpi3B using the general purpose input output (GPIO) pins of the rpi3B.\\
In the following subsections some of the details of the data acquisition system are explained.

\subsection{Hardware} \label{sec:Hardware}
\begin{figure*}[ht]
	\centering
	\includegraphics[trim=0.5cm 1cm 0.5cm 0.5cm, width=1\textwidth]{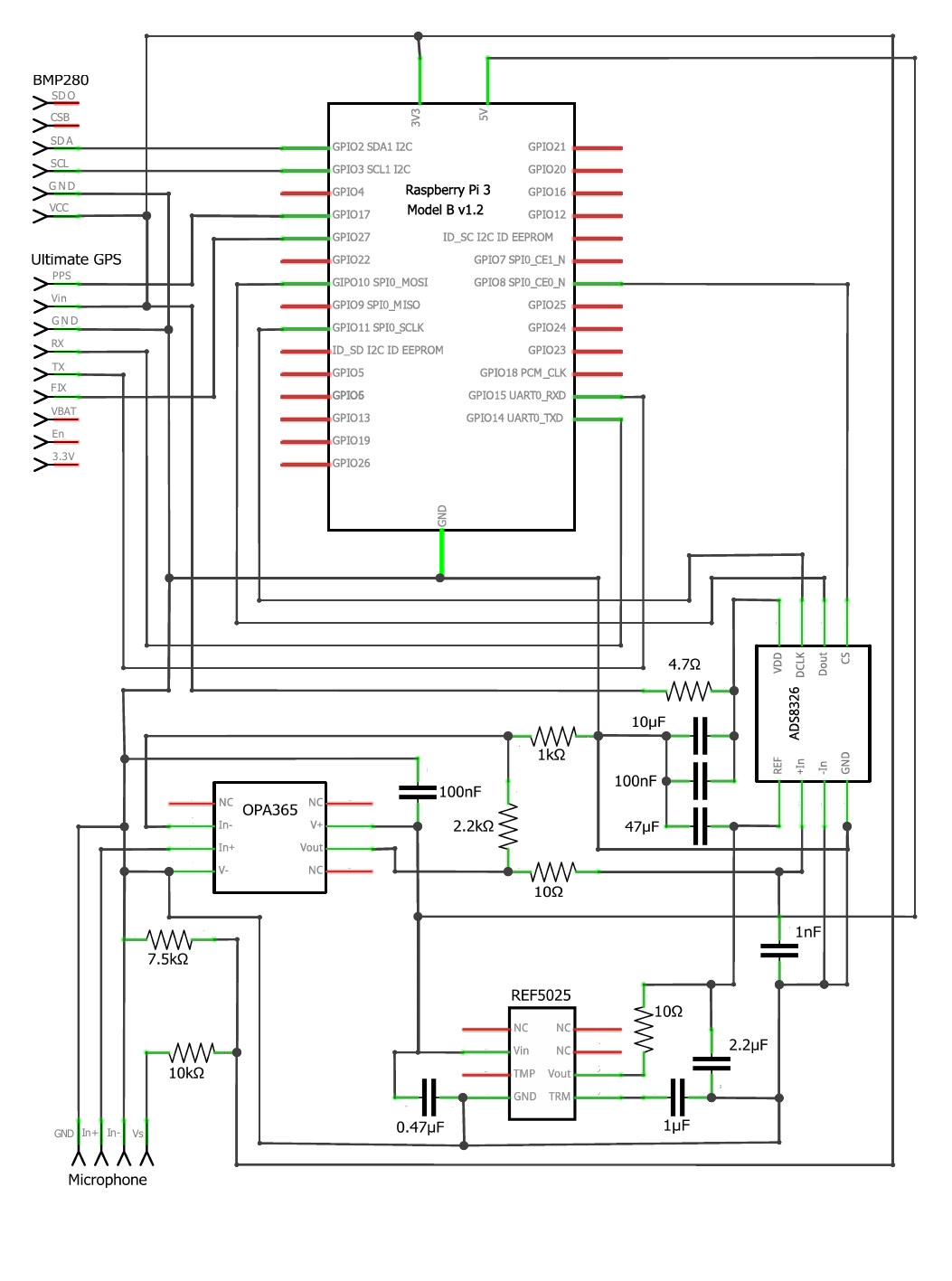}
	\caption{Circuit diagram of the designed PCB}
	\label{fig:PCBdiag}
\end{figure*}

\begin{figure*}[ht]
	\begin{subfigure}[c]{0.346\textwidth}
		\centering
		\includegraphics[width=1\textwidth]{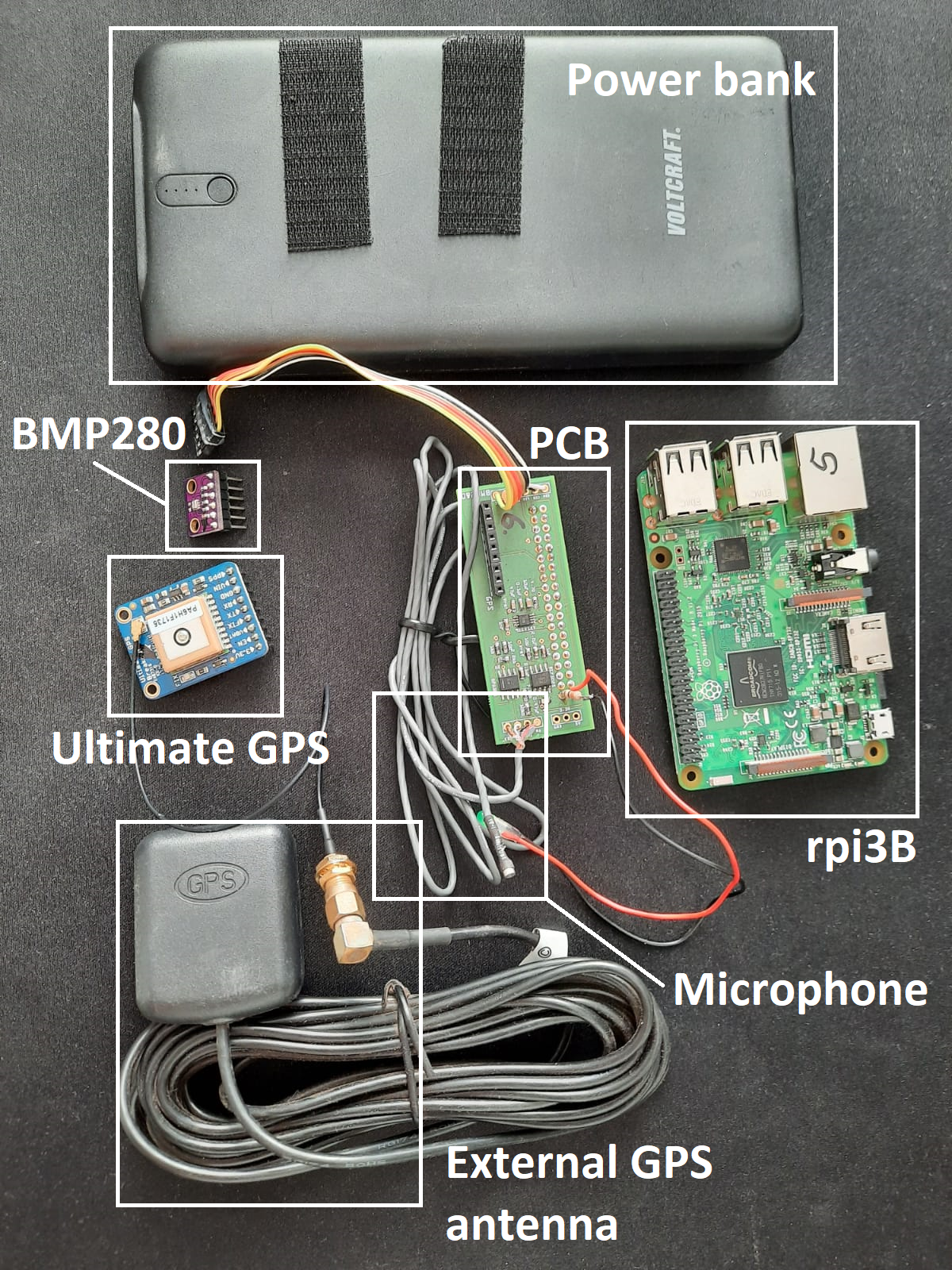}
	\end{subfigure}
	\begin{subfigure}[c]{0.65\textwidth}
		\centering
		\includegraphics[width=0.95\textwidth]{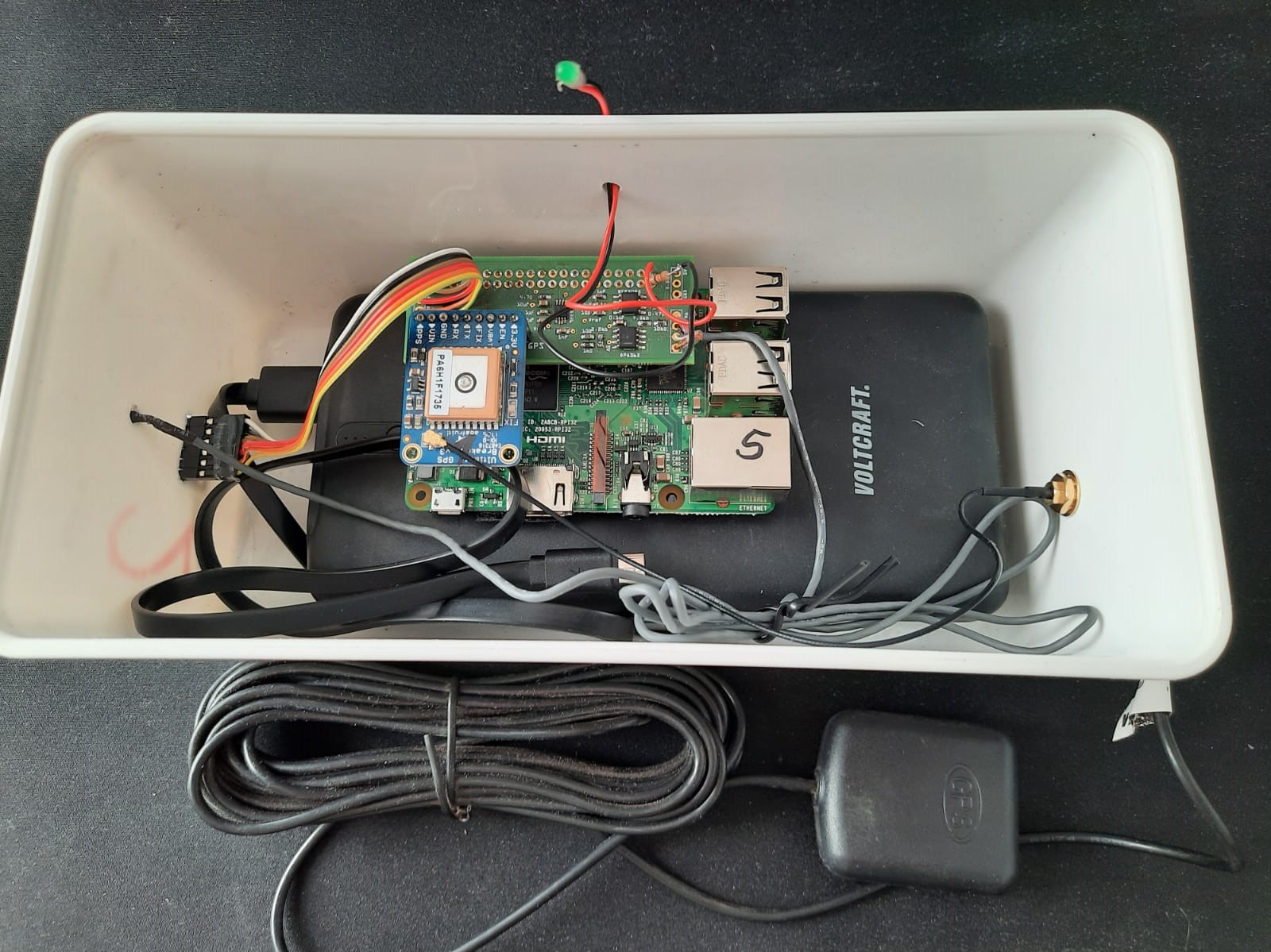}
	\end{subfigure}\\
	\begin{subfigure}[c]{1\textwidth}
		\centering
		\includegraphics[width=0.8\textwidth]{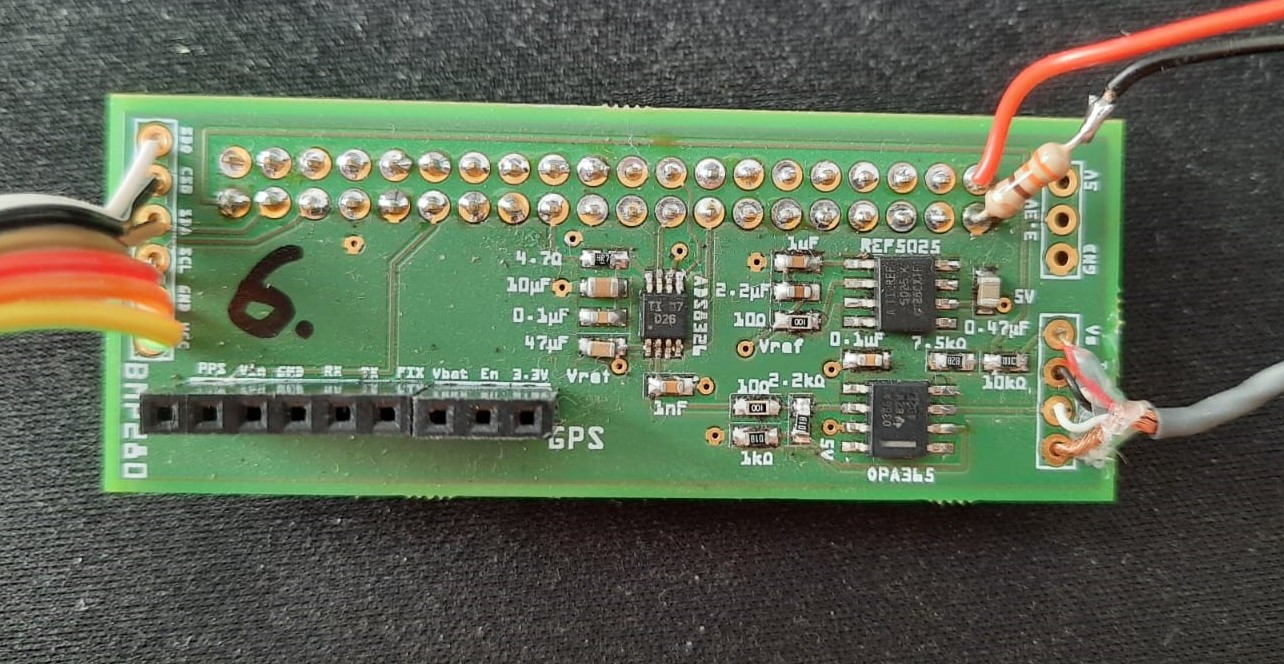}
	\end{subfigure}
	\caption{Parts of the module, assembled version and PCB details.}
	\label{fig:moduleParts}
\end{figure*}

Besides the rpi3B, which acts as the central control unit of each module, additional components are used to build one of the data acquisition modules. For the ADC a Printed Circuit Board (PCB) was designed to connect to the GPIO pins. Besides the actual ADC chip, it contains additional parts needed to run the ADC chip in a data acquisition setup, like a reference voltage, an operational amplifier. The PCB also has connector pins for a GPS module that is needed to record the position as well as the PPS signal and connector pins for an optional sensor to measure temperature and ambient pressure. Figure \ref{fig:PCBdiag} shows the circuit diagram of the designed PCB, where all the necessary parts can be found as well. The used ADC is the model ADS8326, the reference voltage is the REF5025 and the operational amplifier is the OPA365, all from Texas Instruments. The ADC has a maximum sample rate of 100~\si{\kilo\Hz} with an accuracy of 16~bit. The ADC interfaces with the rpi3B via the Serial Peripheral Interface (SPI) protocol. The supply voltage for the ADC is provided via the reference voltage, which in turn is supplied from one of the 5V pins on the GPIO pins of the rpi3B. The microphone sensor is the Knowels FG-23329-P07. It's signal is amplified through the operational amplifier before it is routed into the ADC. The GPS receiver is the Ultimate GPS module from Adafruit. Its communication with the rpi3B is realized via a serial interface. To improve the GPS signal quality and thus get a more accurate position from the GPS, an external GPS antenna was used on each module. The GPS has a raw localisation accuracy of 3~\si{\m}. The optional part to measure temperature and ambient pressure is the BMP280. Its interface to the rpi3B uses the Inter-Integrated Circuit (I2C) protocol. We use a variant that comes on a integrated circuit (IC) chip which can be ordered from various suppliers.\\
The collected data is stored on a USB memory drive that connects to on fo the USB ports on the rpi3B. Since the rpi3B needs a supply voltage of 5~\si{\V}, a simple power bank can be used to power each module. A power Bank with a capacity of 20000~\si{\milli\A\hour} is sufficient to power a module for more than 30 hours. For this prototype, all the parts are assembled into a box for the field operation, which has small holes cut in for the microphone sensor and (if needed) for the BMP280 as well as for the connector to the external GPS antenna. Those holes are then sealed using glue from the inside with the according parts in place. To reduce noise from wind and provide some protection against weather, the microphone sensor and the BMP280 is covered with a wind screen. In this setup, the modules are weather proof for light rain. Figure \ref{fig:moduleParts} shows all the parts of the finished module (disassembled and assembled). The small LED shown in the picture has no important function other than to signal if the module is running.\\
The effort to build one module is minimal. Especially if the PCB can be ordered from a manufacturer that equips it with all the electrical parts. In this case all the parts only need to be connected and fitted into the boxes/casings. But also when the electrical parts still need to soldered onto the PCB, the time requirement is only around 40 minutes. To solder the parts, the use of a hot air soldering station is highly recommended. The price of all parts lies around 200€ to 250€, depending on availability and the used service to manufacture the custom PCB.\\

\subsection{Software}
As mentioned before, the linux based OS (raspbian) that normally runs on the raspberry pi is not suitable to implement a data acquisition system with accurate sample rates in the order of magnitude of 10~\si{\kHz} to 100~\si{\kHz}. The process scheduling, not designed for real time applications, impedes strict timing. Instead of using a real time variant, a custom bare metal OS was implemented. This way the timings and functions can be controlled at the lowest level. When done correctly, a real time OS can be implemented for the rpi3B, where the exact timing is essential for the data collection. The tool that was used to build the custom OS is ultibo \cite{ultibo}. Ultibo is a bare metal development environment for the Raspberry Pi platform. It implements all the low level functions of the raspberry pi in Free Pascal, making it possible to design the OS without the need to write assembly code. Here some of the specifics will be highlighted on a higher level. For more details, we refer the interested reader to the source code at \url{https://github.com/stebuechho/rPi_DAQ}.\\
Keeping the sample rate of the acoustic signal accurate is the most important requirement of the custom OS. To ensure that, one of the four physical cores of the cpu of the rpi3B is dedicated to only measuring the time and to collect a sample from the ADC at the correct times. This one core only runs this one thread and all interrupts are disabled within the thread. The thread only consists of an infinite loop that monitores an internal clock and collects a sample from the ADC whenever necessary. The collected sample is pushed to one of a fixed number of buffers that all are located in the RAM, each with a fixed length. When one buffer is full, it switches to the next one. The full buffers are queued to be flushed to the USB memory, after which the buffer is emptied and put back into the cycle. The writing of data to the USB memory drive is handled by one of the other cores in a different thread, since exact timing is no requirement here.\\
In parallel to the acoustic signal of the microphone sensor, the PPS signal that the GPS module receives is recorded. The PPS signal is a signal that is accurate down to a few nanoseconds. This makes it very suitable to be a base to synchronize signals from different modules.  Recording the PPS signal in sync with the acoustic signal allows us to synchronize the signals of an arbitrary number of modules in post processing without much effort. This PPS signal is a simple 1 bit signal. Once a second it goes high (1) for a few micro-seconds, while it's low (0) otherwise. The state of this signal at the time of sample collection is simply written to a different field of the same buffer as the acoustic data. The PPS signal is routed out of the GPS module via one of its connector pins to one of the GPIO pins of the rpi3B. In a separate thread, an interrupt is set up that is triggered whenever the input to this pin goes to high, increasing a global counter that counts the received PPS pulses so far. This global counter is checked by the dedicated cpu that collects the data from the ADC. Each time it goes up, the PPS state that is recorded with that sample is 1, while it's 0 if the counter didn't increase in the time between two consecutive data samplings. In addition to the PPS state, the time $dt$ that passed since last data collection in \si{\micro\s} is recorded as well. As the PPS state is only one bit and the smallest data type in Free Pascal is 8 bit large, this can be done without using any extra storage space. The PPS state and the $dt$ information are both encoded within the same 8 bit data point. The PPS state is encoded in the most significant bit, while $dt$ is encoded in the first 7 least significant bits. Therefore, the lowest usable sampling rate to still have correct $dt$ information encoded in the data is $\frac{10^6}{2^7 - 1~\si{\micro\s}} = 7874~\si{\Hz}$. Using lower sampling rates will still work without any issues except for the fact that the $dt$ information is not usable directly. The $dt$ information allows to control if the sampling rate stays accurate over time. The GPS position, including latitude, longitude, and altitude, is also logged hourly to a text file. Here, no buffer system was used and the GPS position that is read from the GPS module via a serial connection is simply written to a text file directly once it was read. The time reference that is needed to assign the recorded PPS pulses to the correct seconds for each module is taken from the GPS time. It is taken with the first recorded PPS and saved in the name tag of the data files.\\
The sampling of the temperature and pressure signal from the BMP280 is carried out on yet another thread. This thread also does not run on the dedicated cpu, since the low sampling rate does not require such accurate timings. It is also written to a buffer system like the acoustic data and PPS data and flushed to the USB memory drive. Sampling of temperature and ambient pressure data is possible with a sampling rate of up to 160~\si{\Hz}. To be able to synchronize the data from both files in post processing, the PPS signal state at time of sampling is also recorded and written in conjunction with the temperature and ambient pressure data.
All the data (microphone data, PPS data and temperature and ambient pressure data) are written as binary data to four different files. One for the acoustic data, one for the PPS signal that is in sync with the acoustic data, as well as one file for the temperature and pressure data (it is encoded in single data field by the BMP280) and one corresponding file for the PPS data in sync with the BMP280 data (if recorded). All the data is recorded as the raw, digital data the sensors/ADC give as output. Conversion to the actual physical quantities is done in data post processing.\\
Important settings like sample rate of the acoustic data and temperature/ambient pressure data as well as the rate with which the GPS position is recorded are controlled via a small configuration .ini-file that simply can be placed onto the USB memory drive before the module is powered up.

\subsection{Calibration}
Since the data acquisition modules record the acoustic signal as the digital data the ADC puts out (in this case, an integer number between 0 and $2^{16} - 1$), a reference or calibration measurement is necessary to be able to convert the digital signal to an acoustic signal in \si{\pascal}. We use a class 1 microphone calibrator that operates at 1000~\si{\Hz} with 114~\si{\dB} to acquire the calibration measurement for each microphone. Using these recorded calibration signals, we compute the gain for each microphone so that the calibration signal matches the 114~\si{\dB}  that the calibrator puts out. The computed gains or calibration factors are saved to a file to be used later in data analysis. Here, a linear transfer function for the microphone sensor is assumed. Since calibrating on the summit is not practical, the calibration measurements were taken on ground level in quiet conditions on the day after the measurements were conducted. 

\section{Experimental Setup}\label{sec:ExpSetup}

\begin{figure*}[ht]
	\begin{subfigure}[c]{0.5\textwidth}
		\centering
		\includegraphics[width=1\textwidth]{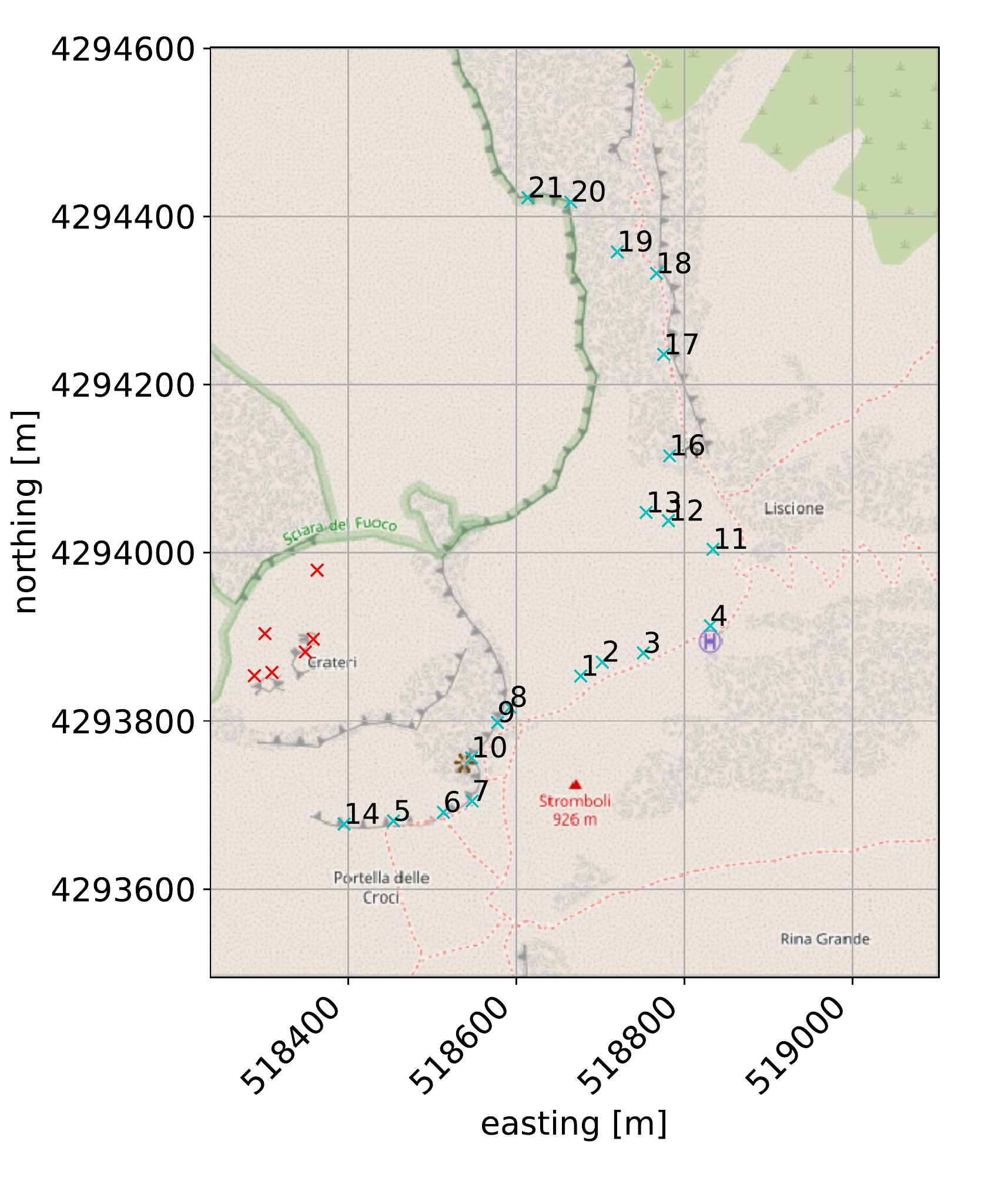}
	\end{subfigure}
	\begin{subfigure}[c]{0.5\textwidth}
	\centering
	\includegraphics[width=1\textwidth]{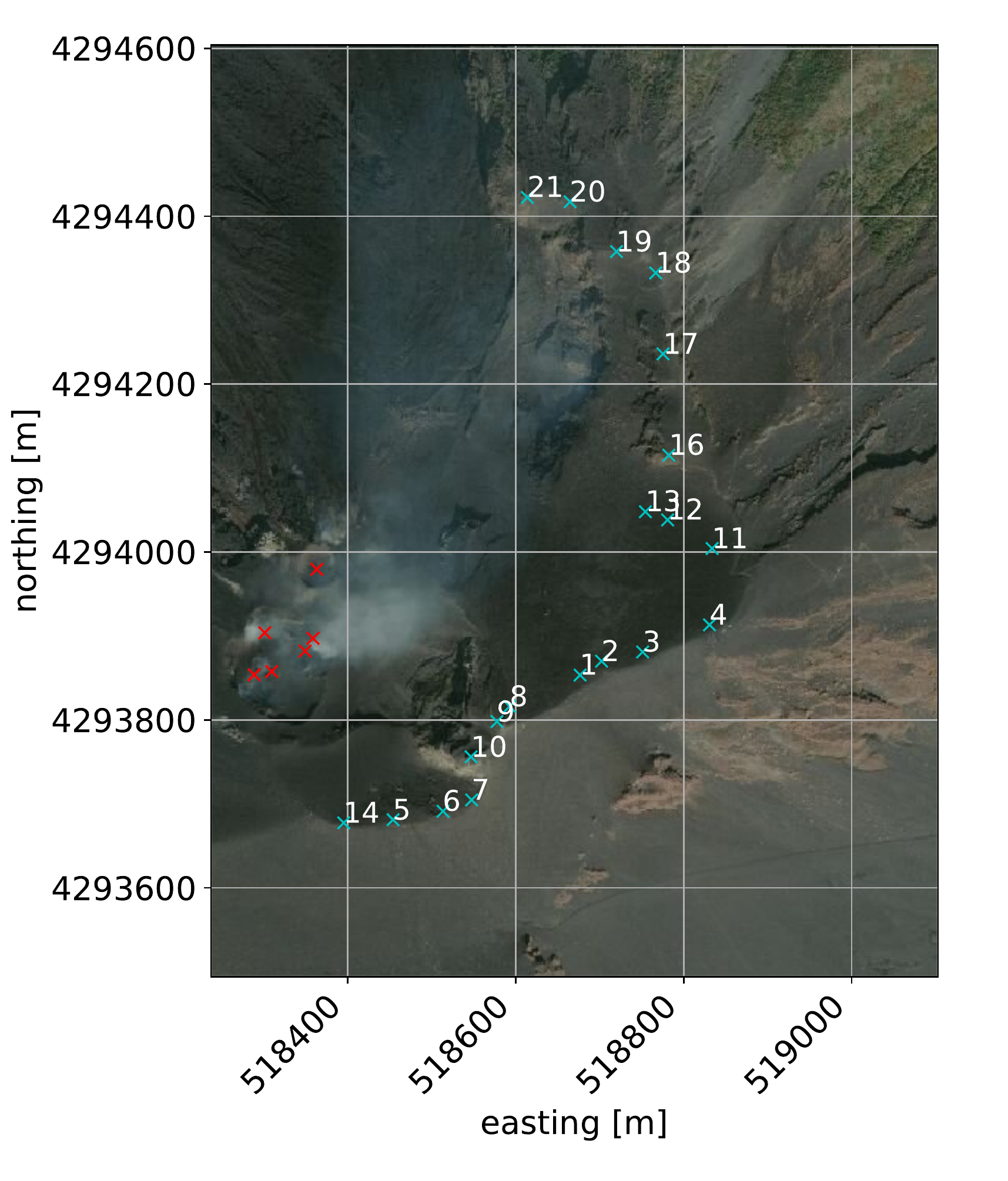}
	\end{subfigure}
	\caption{Positions of the modules around the summit of the volcano. Background maps: OpenStreetMap (left) and EsriSat (right). Axes are in UTM coordinates (zone 33S). Cyan X's mark the positions of the modules, where the white numbers next to the markers show how the modules were enumerated. The red red X's mark the positions of the different vents in the crater on the day of measurement. The locations were determined from drone material \cite{Schmid:2020} that mapped the crater area on the day of measurement. Note: Because of the volcanic activity the crater area is subject to constant change. Therefore the map material shown here is not up to date. }
	\label{fig:micPos}
\end{figure*}

A set of 20 data acquisition modules was deployed on the active volcano \textit{Stromboli} in Sicily, Italy, within the Broadband Acquisition and Imaging Operation (BAcIO) 2019. On May 15th 2019, the data was collected in a time frame from 10h to 16h UTC. The modules where spread over a wide area on the summit of the volcano. Figure \ref{fig:micPos} shows their location on the volcano. All the modules were recording with a sample rate of 31250~\si{\Hz}. During the time of measuring, multiple eruptions occurred from different vents in the crater area. The weather on the day was mixed. It was windy at times, with gusts of wind mainly blowing in the north-west direction. We do not have any information on the wind speed however. Also it was cloudy and the visibility on the crater area was poor. The data is available on request from the authors.\\
During the measurement campaign, a minor problem came to light. When connecting the BMP280 sensor to the PCB, the supply voltage for the microphone sensor is increased to a too high level, leading to signal clipping as the signal to the ADC has a too high gain. The reason could not have been investigated yet, but it is likely that a resistor on the PCB is bypassed on the integrated circuit on the BMP280 chip we used. The resistor that seems to be bypassed is the 7.5~\si{\kohm} resistor close to the microphone connection in figure \ref{fig:PCBdiag}, which is part of the voltage divider (that simply consists of two resistors). The voltage divider is used to decrease the supply voltage to the microphone sensor from 3.3~\si{\V} to 1.4~\si{\V}, as the supply voltage range for the Knowels FG-23329-P07 microphone sensor is between 1~\si{\V} and 1.65~\si{\V}. Therefore the measurements were conducted without the optional BMP280, so no temperature or ambient pressure data is available for the data analysis here, unfortunately. A possible fix could be to simply use another reference voltage IC to deliver the appropriate supply voltage to the microphone sensor instead of a voltage divider. For the sake of completeness, in the following, we treat the setup as if this problem was already fixed and a BMP280 could still be used.

\section{Data Analysis}
\begin{figure*}[ht]
	\begin{subfigure}[c]{0.5\textwidth}
		\centering
		\includegraphics[width=1\textwidth]{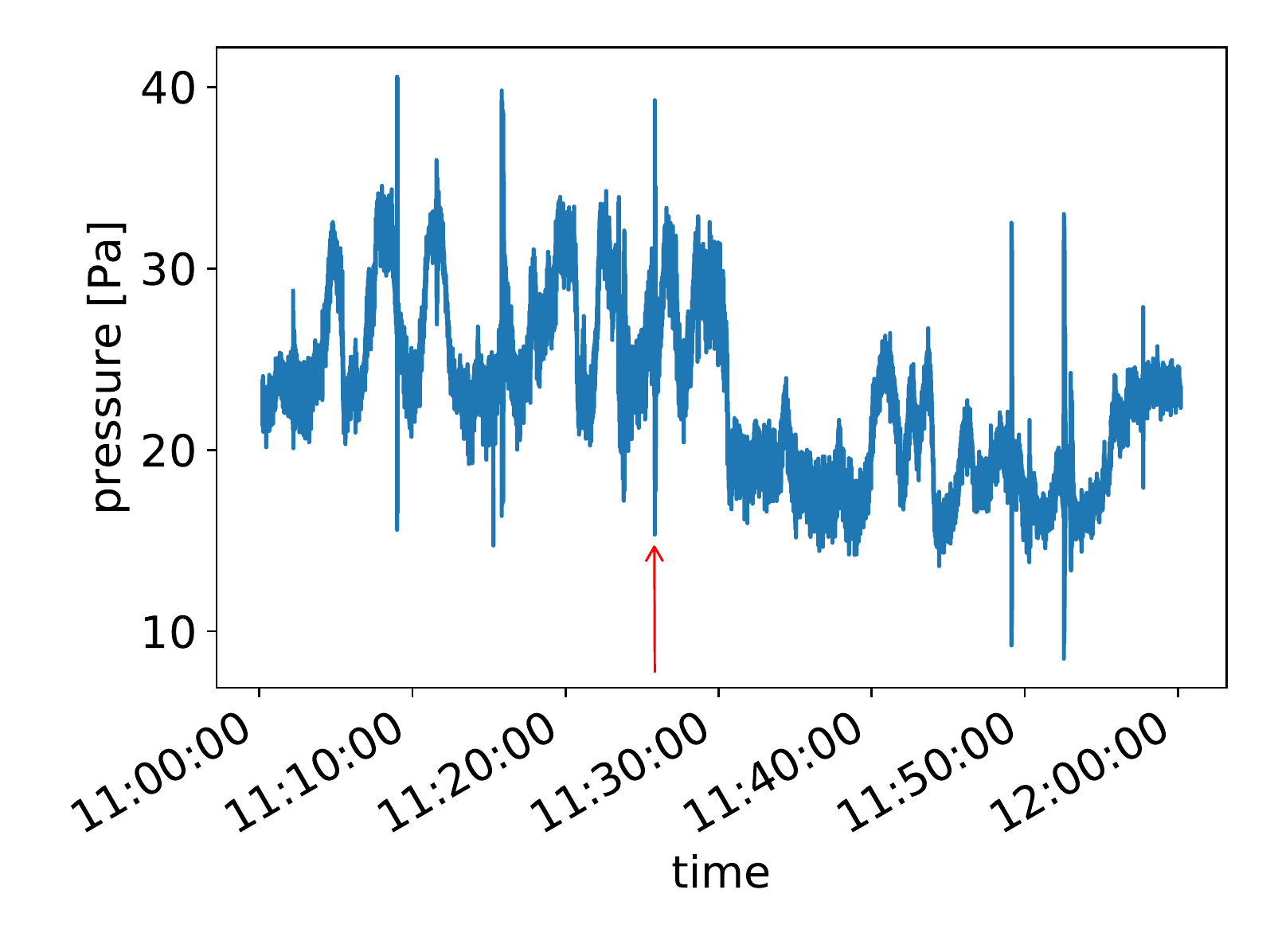}
	\end{subfigure}
	\begin{subfigure}[c]{0.5\textwidth}
		\centering
		\includegraphics[width=1\textwidth]{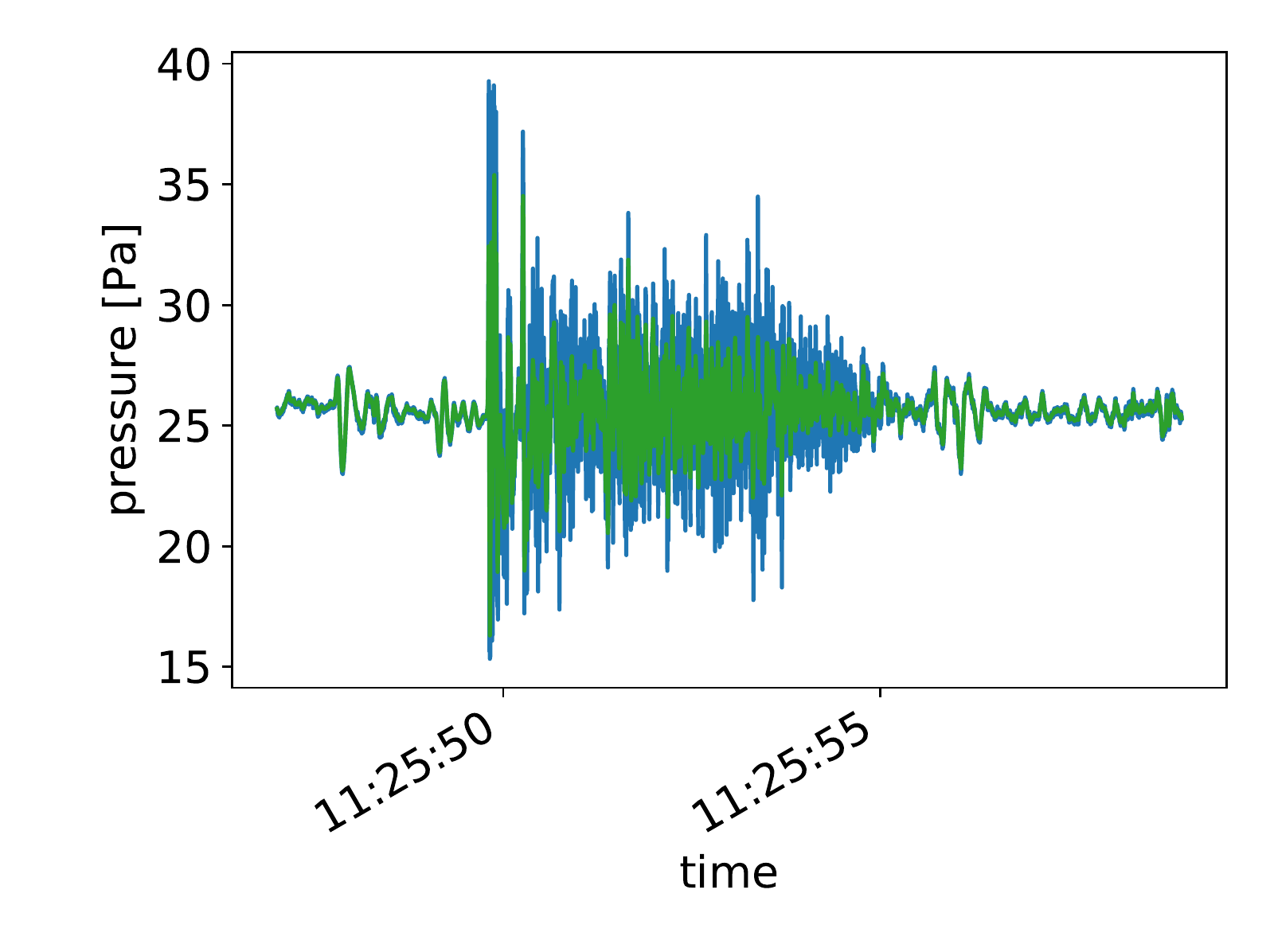}
	\end{subfigure}\\
	\begin{subfigure}[c]{0.5\textwidth}
	\centering
	\includegraphics[width=1\textwidth]{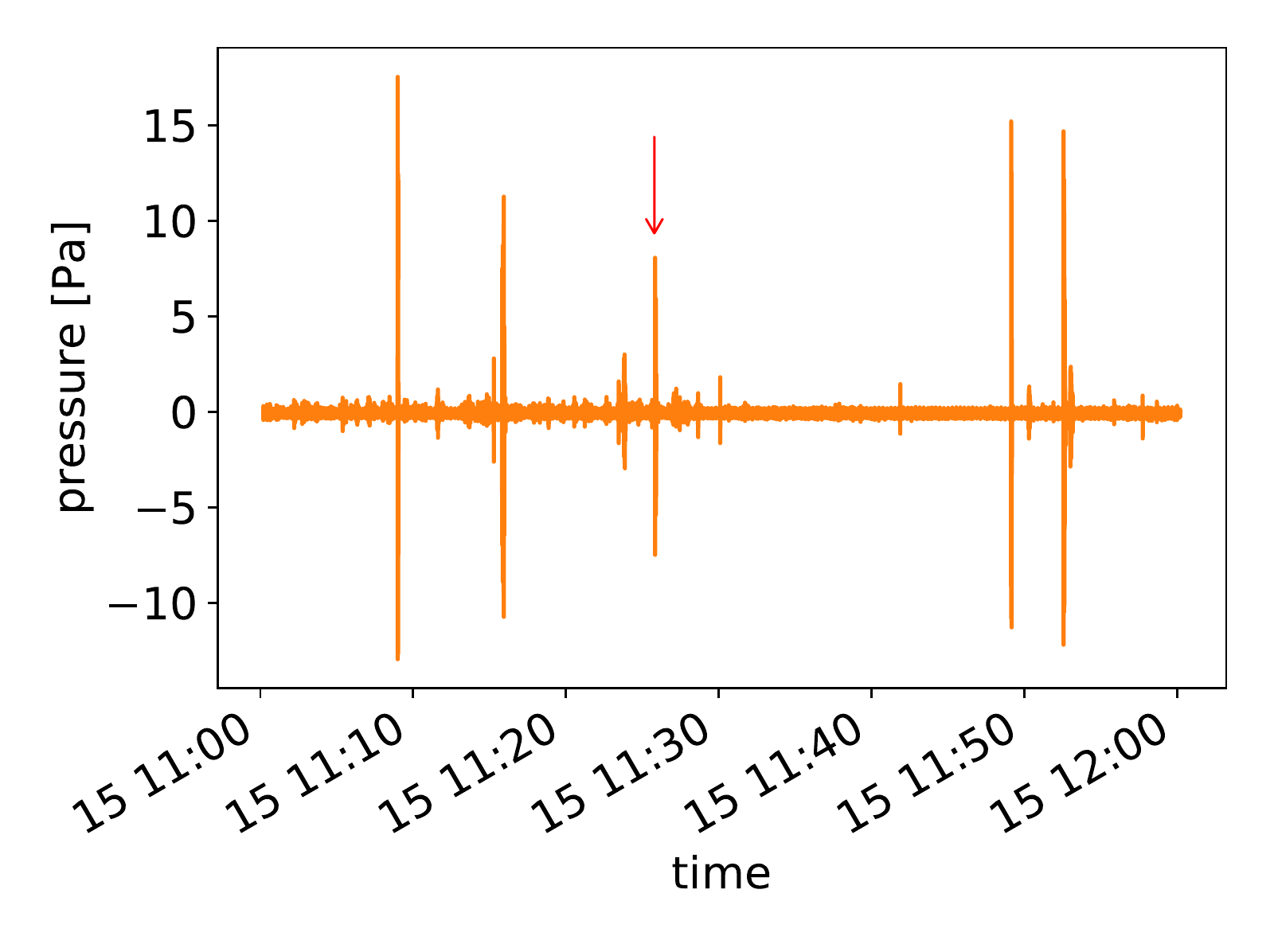}
	\end{subfigure}
	\begin{subfigure}[c]{0.5\textwidth}
	\centering
	\includegraphics[width=1\textwidth]{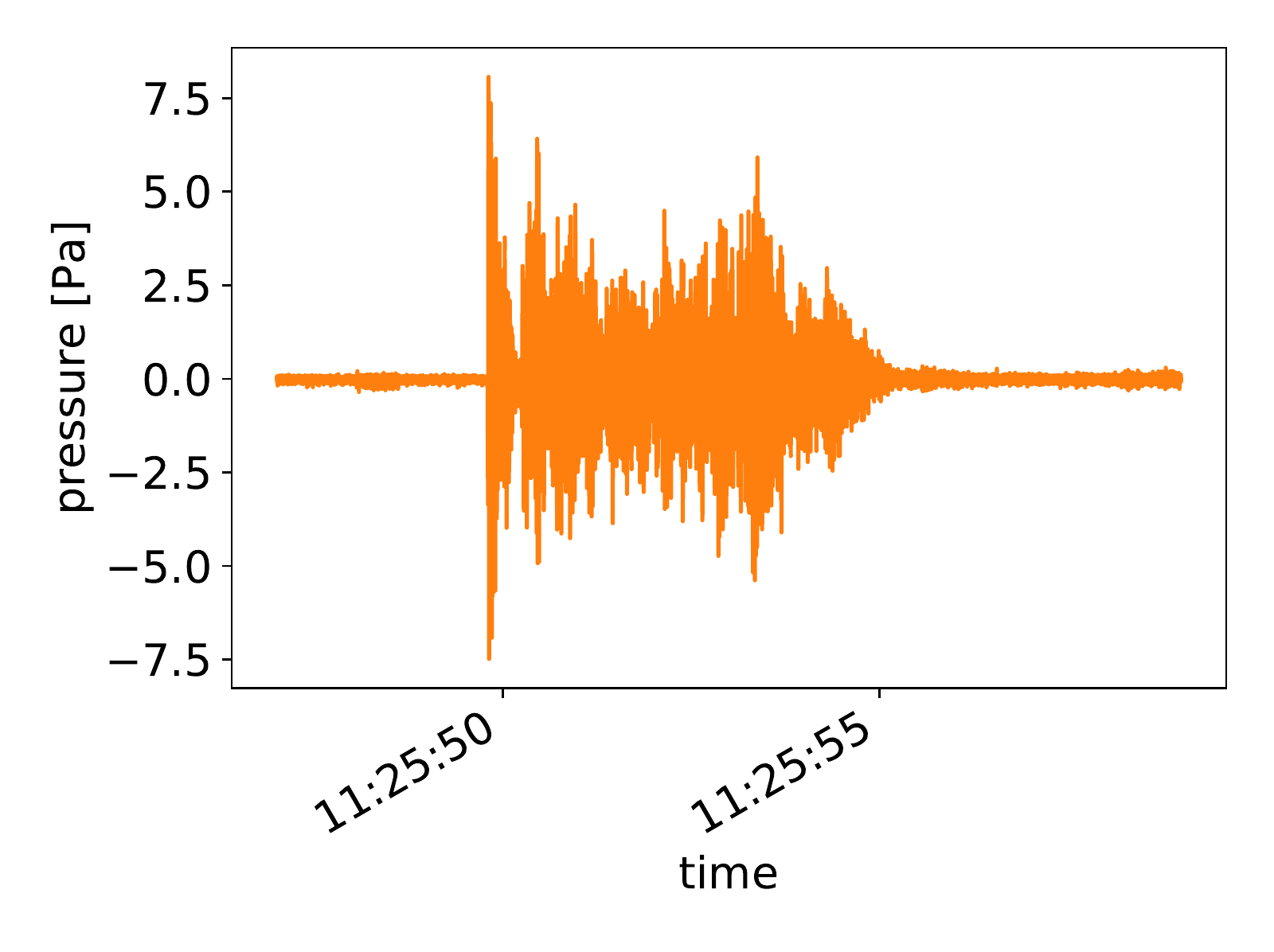}
	\end{subfigure}
	\caption{Example signals from module 5. Top row, left: raw signal as recorded by the module after applying the calibration factor. Top row, right: zoomed in raw signal of one eruption event (blue line) together with the 0.015 sec (481 samples) moving average that is used to filter the raw signal (green line). Bottom row: resulting filtered signals for the same time periods ("blue line minus green line"). The red arrows on the left column mark event that is shown in the right column}
	\label{fig:prefilter}
\end{figure*}

The data is stored in it's raw digital form, as mentioned above. Applying the gain factors that were computed from the calibration signals converts the digital microphone signals to pressure data in \si{\pascal}. The microphone sensors that were used in the setup have a lower cut off frequency of around 60~\si{\Hz}. Therefore, we apply a high pass filter with a filter frequency of 65~\si{\Hz}. This is realized in time domain by subtracting the centred moving average over $481$ samples. With the original signal being sampled at 31250~\si{\Hz}, the resulting filter frequency is $\frac{31250~\si{\Hz}}{481} \approx 64.97~\si{\Hz}$. Figure \ref{fig:prefilter} visualizes the different steps on data examples.\\
The data field that contains the PPS and $dt$ information needs to be split back into the corresponding data. This can be done by simple bit operations. Synchronization of the signals of different modules can be easily done by simply aligning the acoustic signals using the PPS pulses that were recorded in parallel. The $dt$ information shows a constant value of 32~\si{\micro\s} corresponds to the sample rate of 31250~\si{\Hz}, across all modules and is not shown here.\\
If recorded, the temperature and ambient pressure data needs to be converted from its digital form the BMP280 sensor puts out as well. Here, a special algorithm needs to be implemented that uses some calibration information stored in permanent memory on each of the sensors. This data is read from the BMP280 sensor during each start up of a module and written to a text file. The algorithm to convert the data is provided within the BMP280 data sheet.
After all the signals are converted and synchronized, the eruptions (events) need to be identified and extracted from data. For each of the detected events, we identify the location of the sound source.\\

\subsection{Event detection - preselection}
\begin{figure*}[ht]
	\begin{subfigure}[c]{0.5\textwidth}
		\centering
		\includegraphics[width=1\textwidth]{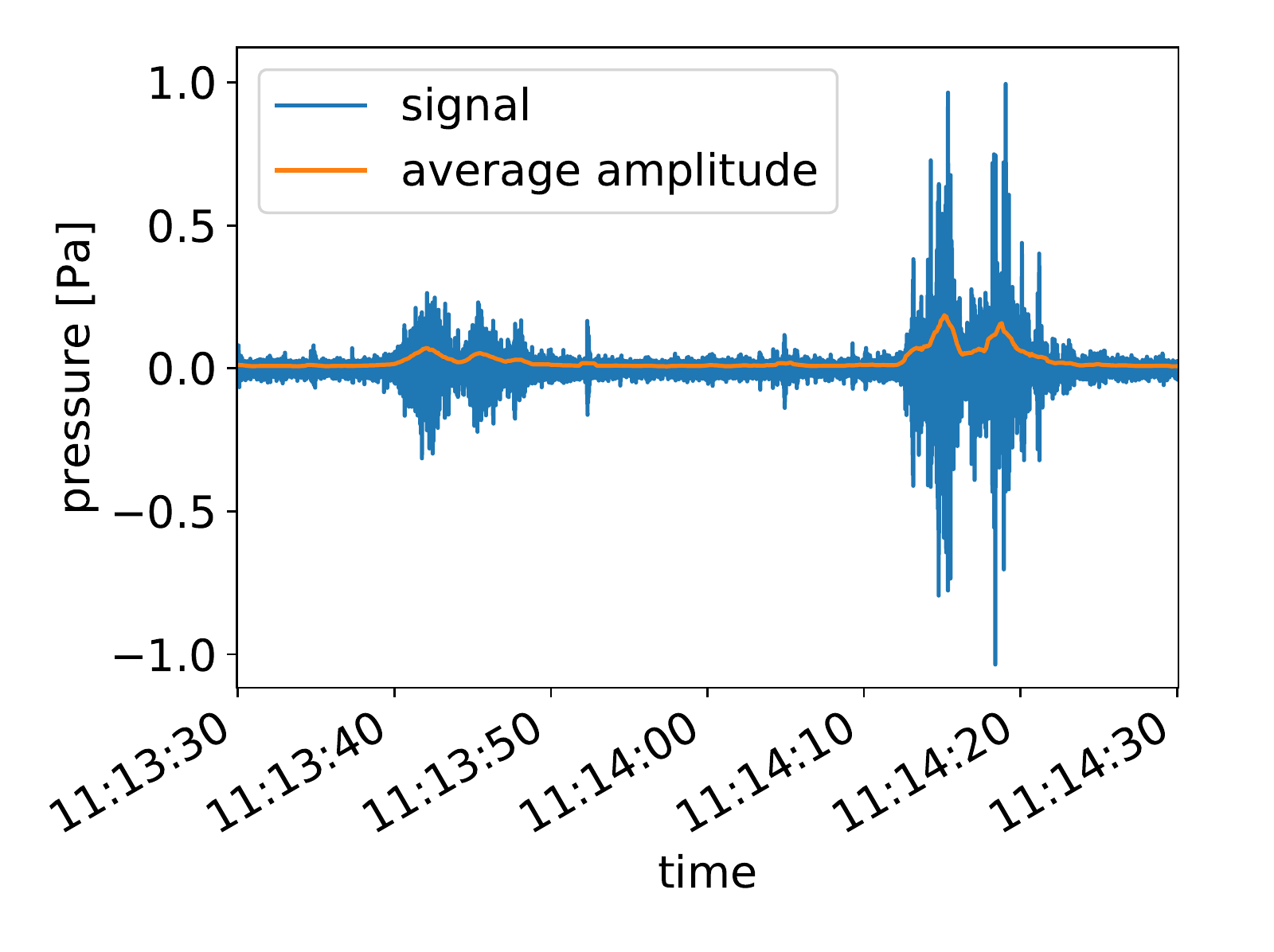}
	\end{subfigure}
	\begin{subfigure}[c]{0.5\textwidth}
		\centering
		\includegraphics[width=1\textwidth]{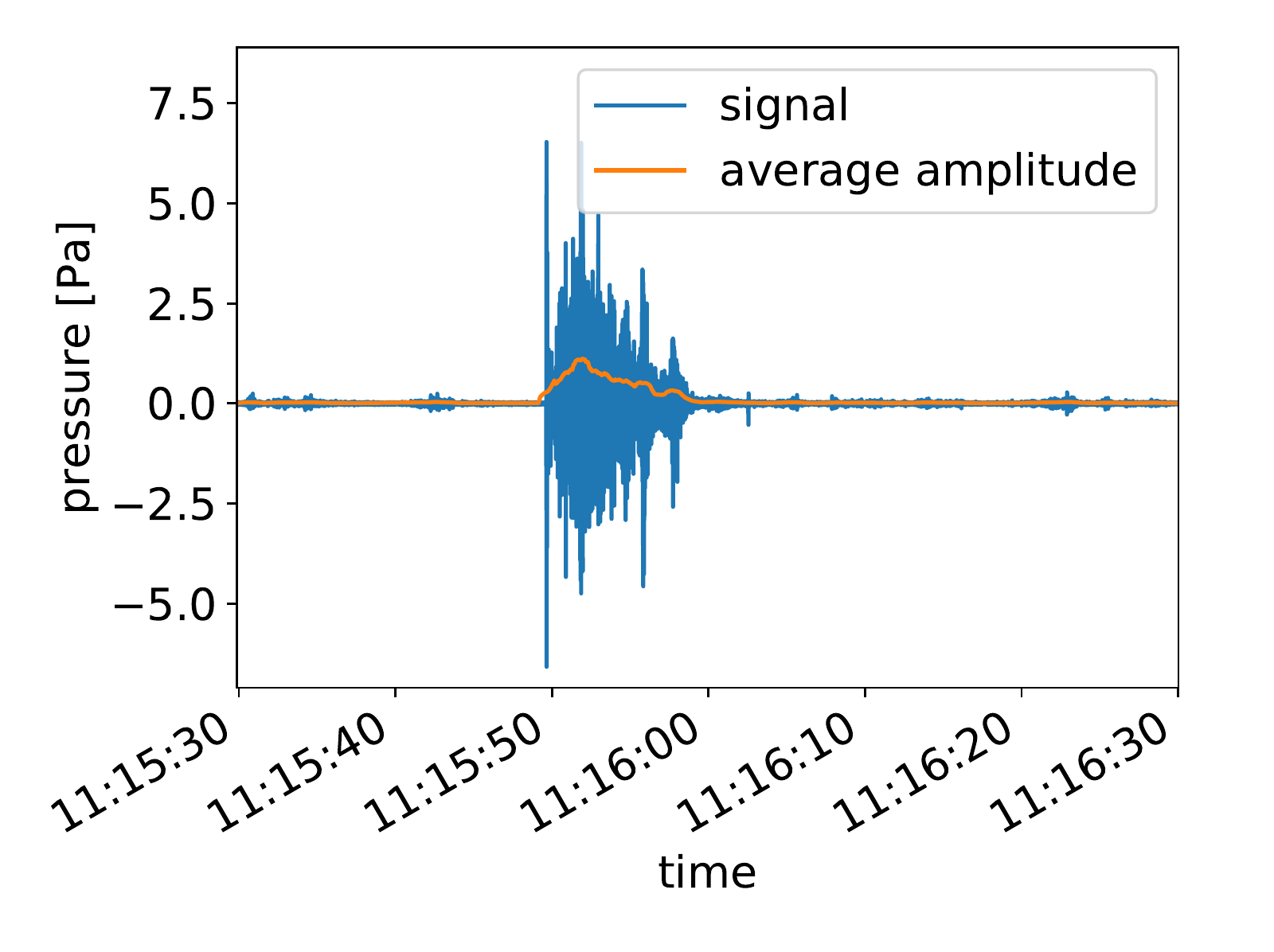}
	\end{subfigure}\\
	\begin{subfigure}[c]{0.5\textwidth}
		\centering
		\includegraphics[width=1\textwidth]{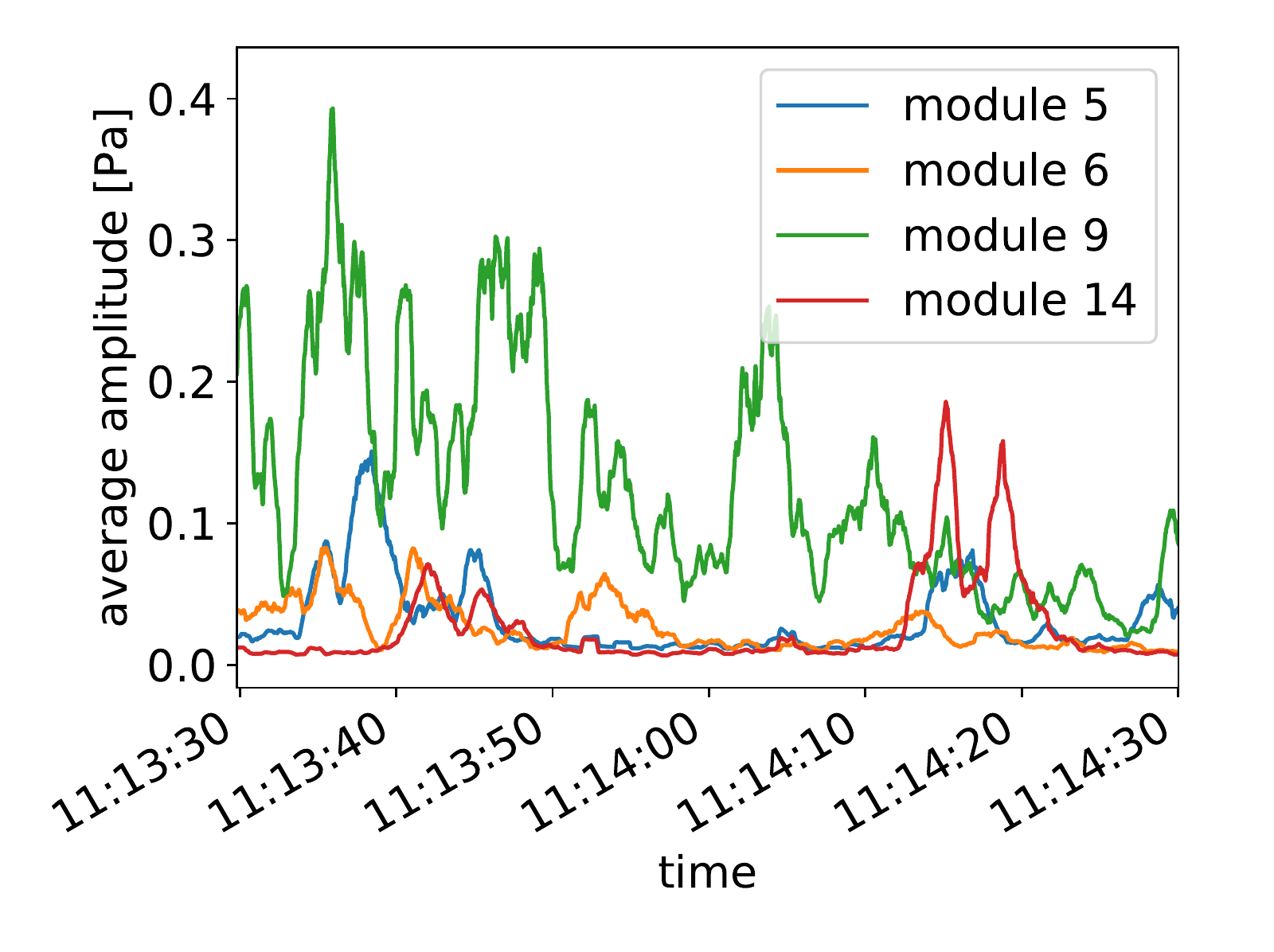}
	\end{subfigure}
	\begin{subfigure}[c]{0.5\textwidth}
		\centering
		\includegraphics[width=1\textwidth]{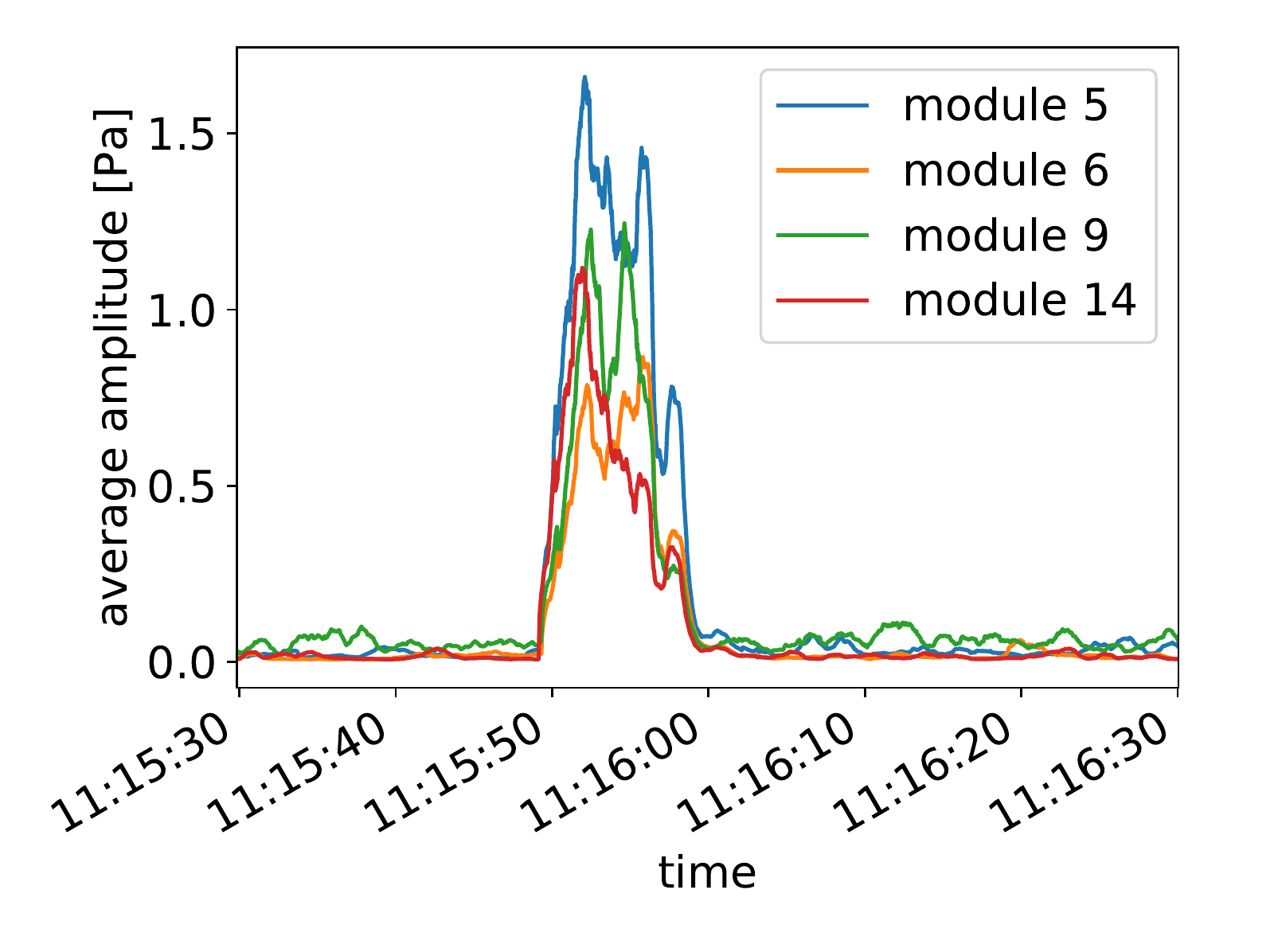}
	\end{subfigure}
	\begin{subfigure}[c]{0.5\textwidth}
		\centering
		\includegraphics[width=1\textwidth]{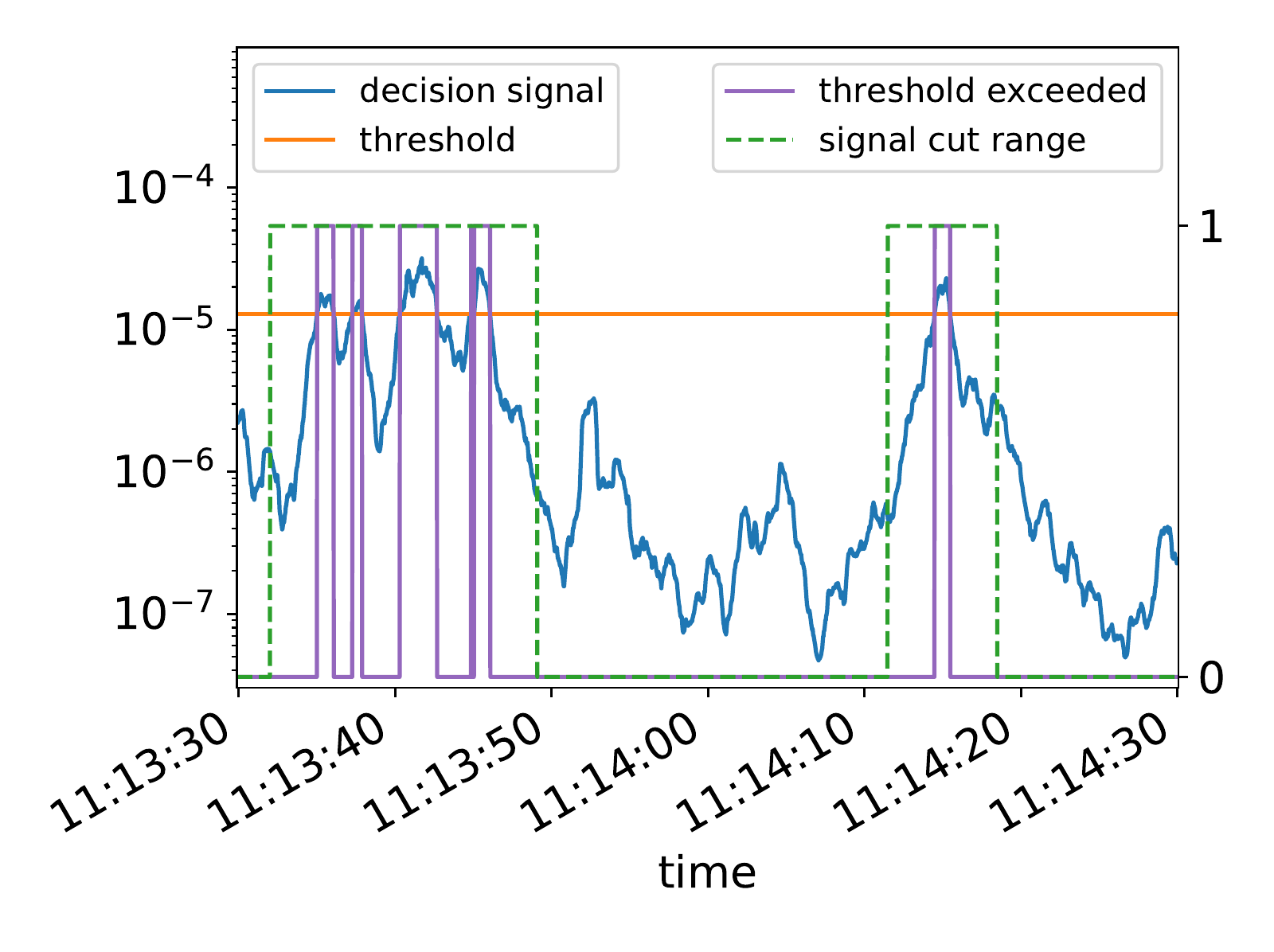}
	\end{subfigure}
	\begin{subfigure}[c]{0.5\textwidth}
		\centering
		\includegraphics[width=1\textwidth]{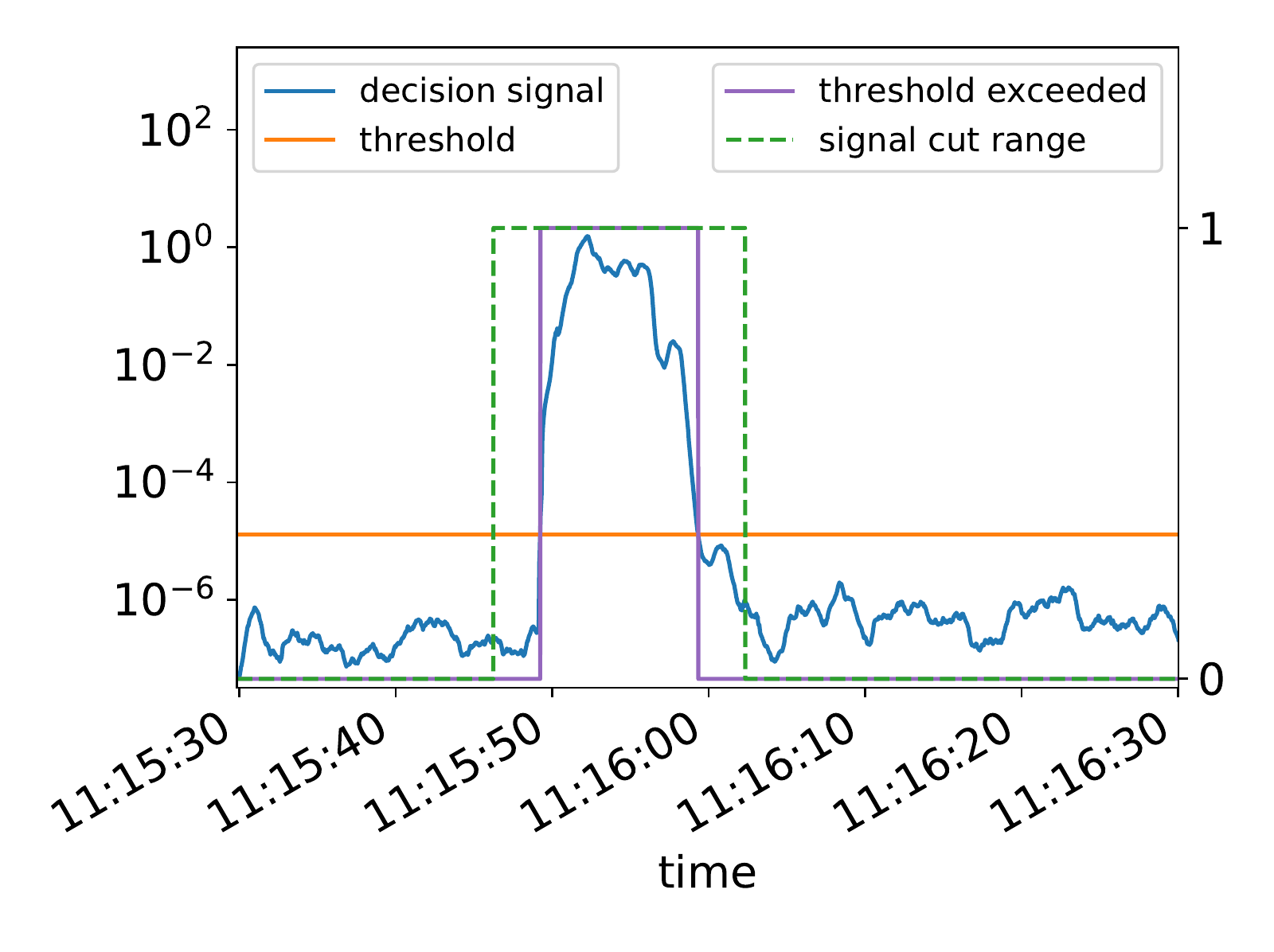}
	\end{subfigure}\\
	\caption{Results of the steps of the event detection procedure. Left column: example of a weak event (most likely noise from wind). Right column: example of a strong event (an actual eruption). Top row: time series of the signal of module 14 (blue) and average amplitude (orange). Middle row: average amplitudes of all modules used for event detection. Bottom row: resulting detection signal (blue) and event detection threshold (orange). Due to the different orders of magnitude the detection signal ranges, both is plotted on a logarithmic axis (left axis). The purple line denotes where the detection signal exceeds the threshold and the dashed green line illustrates the resulting time range that is identified as a single event and saved to a separate file.}
	\label{fig:event_detect_thresh}
\end{figure*}

To identify all the acoustic events in the collected data we use a simple algorithm that analyses the moving averages over the signal amplitudes of the modules close to the crater area (and thereby close to the acoustic events from eruptions). For the results shown here, 4 modules were used (modules 14, 5, 6 and 9). With 4 modules, the chances that all modules pick up correlated noise from a wind gust is small, reducing the number of false detections. Using more modules is possible, but should not yield much better results, as this only serves as a preselection. In the analysis of the events described further below, all modules are considered.\\
To reduce the computational load, we first downsample the 4 time series to a fifth of it's sample rate (6250~\si{\Hz}). Then we compute the centred moving average over a duration of one second of data and take the time steps wise product of the different resulting moving averages of the different modules. The resulting values are compared with a threshold value of $(0.06~\si{\pascal})^n$, where n is number of modules used for event detection ,$n=4$. Each data point in time where that threshold is exceeded is a candidate for being part of an event.
Since the duration of some of the events is quite long (up to 20~\si{\s}) and also because over the duration of an event the sound amplitude can vary a lot, it may happen that during an event, the threshold is not exceeded at each single time step. Therefore, we smooth it by computing yet another centred moving average over the boolean signal that marks the time steps where the threshold is exceeded. This last moving average runs over a time duration of 3~\si{\s}. The result is a signal that is $> 0$ in the range around any potential event, with an extra margin of about 1.5~\si{\s} in the front and the back of the event. Based on this signal, we independently cut each of the recorded signals of all modules into chunks, where each chunk consists of all consecutive data points in time of all modules where the signal is larger than zero. Figure \ref{fig:event_detect_thresh} shows the results of the different steps of procedure on two examples of the recorded data, a very weak event and another example for an event with a larger amplitude. This method does not distinguish between event types. It simply assumes an event when the signal amplitudes of all analysed modules is above a certain threshold at the same time. Because of that, it may detect "false positives", meaning that it may also detect an event when there is any source of sound that is not from an eruption. In this setting, this could be mostly noise from strong winds. This needs to be considered and handled during further analysis of the event signals.
\\

\subsection{Source localization}
\begin{figure*}[ht!]
	\begin{subfigure}[c]{0.5\textwidth}
		\centering
		\includegraphics[width=1\textwidth]{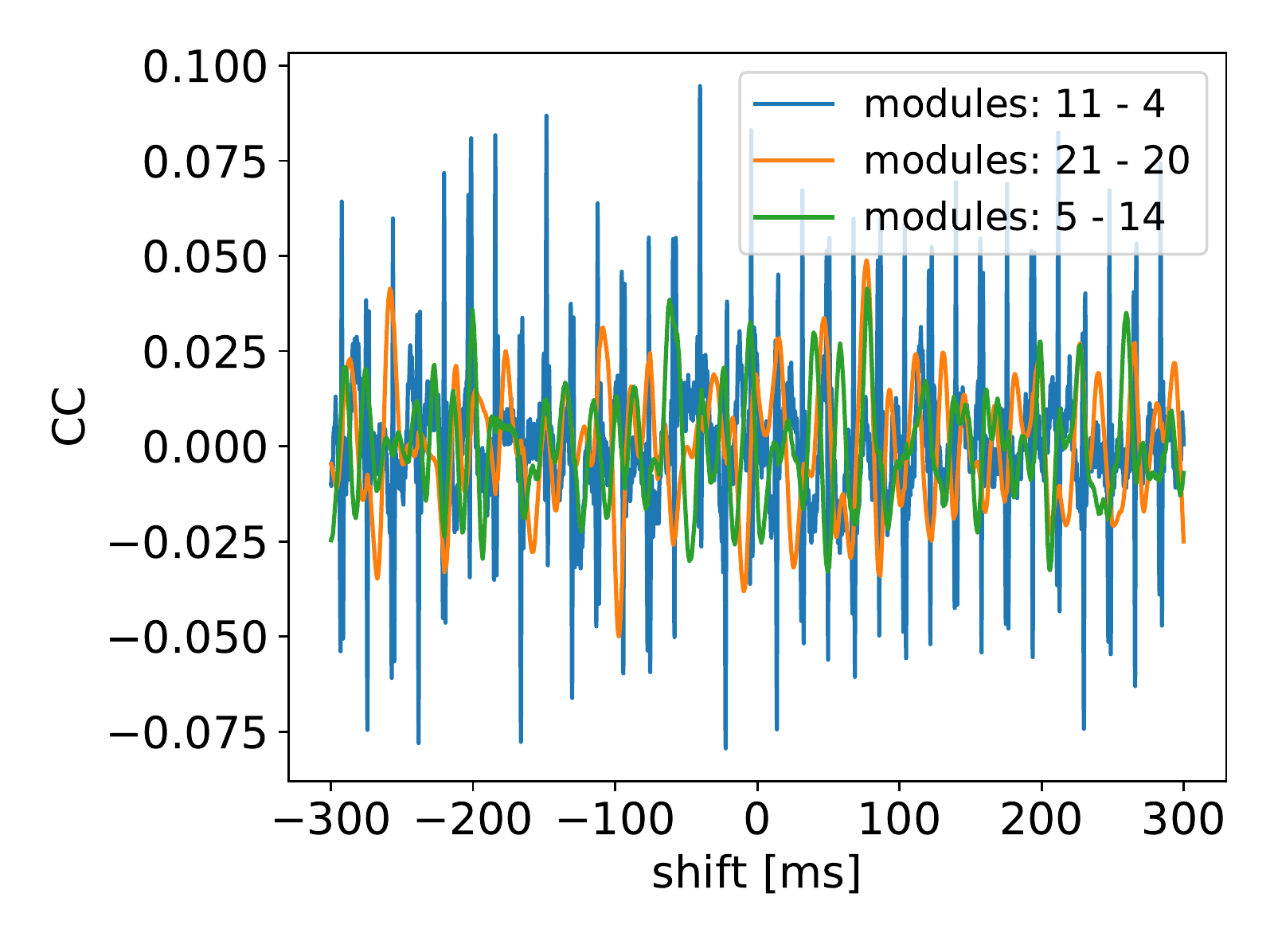}
	\end{subfigure}
	\begin{subfigure}[c]{0.5\textwidth}
		\centering
		\includegraphics[width=1\textwidth]{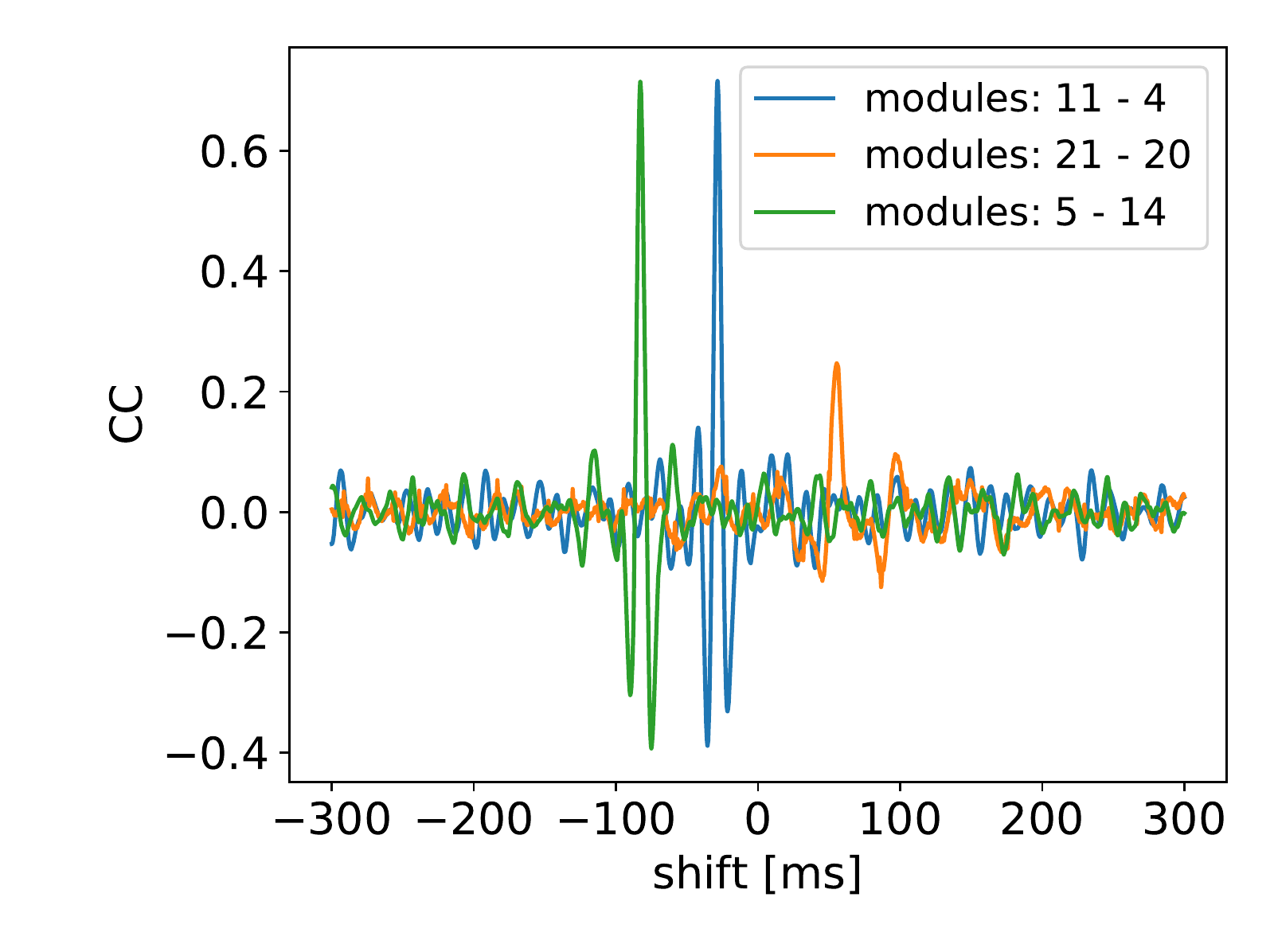}
	\end{subfigure}
	\caption{Cross correlation coefficients for different shifts from two different events for a selection of three different module pairs. Left: weak event (event around 11:13:40 h in figure \ref{fig:event_detect_thresh}, left column), Right: strong event (event around 11:15:55 h in figure \ref{fig:event_detect_thresh}, right column)}
	\label{fig:crossCorrMap}
\end{figure*}

The \textit{Stromboli} has multiple active vents in the crater area (see the red x in figure \ref{fig:micPos}). During the time period the data was recorded, there were 6 different visible vents in the crater area, while not necessarily all were erupting. Since the characteristics of the eruptions can vary strongly between the vents, it is important to be able to distinguish between them in the data. We implement an algorithm based on a Time Difference of Arrival (TDoA) Method \cite{TDoA} to find an estimate of the coordinates where each of the events occurred. TDoA is a common technique to localize the source of a signal from the time difference of arrival of that signal at different spatial locations. Using the speed with which the signal propagates (the speed of sound), the time differences can be converted into differences in distance to the source location. The time difference of arrival between the signals of two modules can be found using the (normalized) cross correlation between these signals. The cross correlation coefficient (CC) of two time discrete signals for a given sample shift $j$ can be computed from:

\begin{equation}\label{eq:cross_cor}
CC(s_1, s_2, j) = \frac{1}{n_1 - n_0}\sum_{n = n_0}^{n_1} \frac{s_{1,n} \cdot s_{2,\ n  + j}}{\sigma(s_1)\cdot\sigma(s_2)}
\end{equation}\\

where $s_{1/2}$ denotes the time discrete signals from two different modules, $n_0$ and $n_1$ denote the start and end sample of the signal range that is to be considered. $\sigma(s)$ is the standard deviation of the signal $s$ on the interval from $n_0$ to $n_1$.\\
As figure \ref{fig:micPos} illustrates, the position of the modules relative to the crater area varies a lot in distance, angle and altitude. Thus, it can be expected that the signal waveforms of the same event look very different between the modules with very different positions. Since we use the cross correlation coefficients to identify the TDoA between signals, it is beneficial when $s_1$ and $s_2$ have similar waveforms. Thus, we take the pairwise cross correlation of two modules that are closest to each other, starting from the module closest to the crater area (module 14). From there we "walk" along from module to module, computing the cross correlation coefficients between the two signals. Meaning we start with the pair of module 14 and 5, go on with the pair of module 5 and 6 and so on, to eventually end at the pair of module 20 and 21. To determine the distance between modules, we simply use the GPS coordinates the modules recorded. More details on that are given in section \ref{sec:GPSacc}. For each pair, we determine the cross correlation coefficients for a range of $j$ samples  that covers approximately 0.3~\si{\s} in both directions. We choose $n_0$ and $n_1$ for each event to range over the whole signal as written out by the event detection algorithm described above. The shift $j$ with the largest correlation coefficient marks the TDoA between the two signals in number of samples, or in seconds when divided by the signals sample rate. Figure \ref{fig:crossCorrMap} shows the cross correlation coefficients of 3 module pairs over different $j$ for 2 different example events. It can be seen that some of the event data of the modules have a poor signal to noise ratio (SNR), especially the modules further away from the events and when the eruption is not very loud. In those cases the signal is not well correlated with the paired module and there is no distinct peak in CC, whereas when the SNR of both signals is high and they are well correlated, there is a very distinct peak in CC and the curve is very smooth. To identify those signals with poor SNR, the ratio between the maximum value of CC and the standard deviation of the high pass filtered CC (as a measure of its "non smoothness") is taken. This ratio is used as a measure of the quality of the CC. When the ratio is high, we have a strong peak and a smooth CC signal. If this ratio is below the value of 8, we consider the two signals to be not correlated. The threshold of 8 was found empirically. In this case we skip the "downstream" module with the poor SNR for the current event and pair the "upstream" again with the next module "in line" (downstream, upstream and in line refers to the order the modules are paired as described above). If more then 11 of the 20 modules are skipped for one event, we discard the whole event and assume the event detection triggered due to loud noise like strong wind and no eruption took place or because the eruption was not loud enough so most of the signals have poor SNR.\\
This way, we acquire the TDoA between all modules for each event, while also filtering out events that are likely no eruptions / are not analysable.\\

Once all TDoA between the signals of the modules for one eruption are determined, the location of the source can be approximated. At this point, we differ from the approach normally used in the TDoA method, which solves a system of hyperbolic equations (one for each module pair) that describes the dependency of the difference in distance on the source location. Since we already know the rough area of the source (the crater area), we simply use a least squares method on a virtual grid that discretizes the crater area. On a high level, we estimate the source locations by comparing the set of TDoA we found from analysing the signals of the different modules with the TDoA that would result from different test source locations. In detail, as mentioned above, for the test locations a 3D grid that covers all of summit is set up. On this grid, we compute the distance of every grid point to each module. With the strong assumption of a constant speed of sound of 343~\si{\m\per\s} and no wind influencing the sound propagation, the distance is transformed into a TDoA by multiplying with the speed of sound. With this, we have a 3D map of the perimeter around the summit of the volcano and for each point we know the time a signal needs to travel to any of the 20 modules (given the assumption of a constant speed of sound and no wind influence) forming a travel time vector over the 20 modules in each grid point. Subtracting the smallest value in each travel time vector from each travel time vector, gives the TDoA of all modules to the module where the signal arrives first, in all grid points. With this, a localization error for each grid point is calculated, by taking the mean squared error of this TDoA vector in every grid points with the vector of TDoA determined from cross correlating the signals of the modules for each event. The grid point with the smallest mean squared error between "signal TDoA" and "grid TDoA" is taken as the estimate of the location of the sound source. In order to get a accurate estimate, we start with a resolution of 5~\si{\m} to set up the grid around the crater area, and then make a second pass with a resolution of 1~\si{\m} around each of identified source positions.\\


\section{Results}\label{sec:Results}
\begin{figure*}[ht]
	\centering
	\includegraphics[width=1\textwidth]{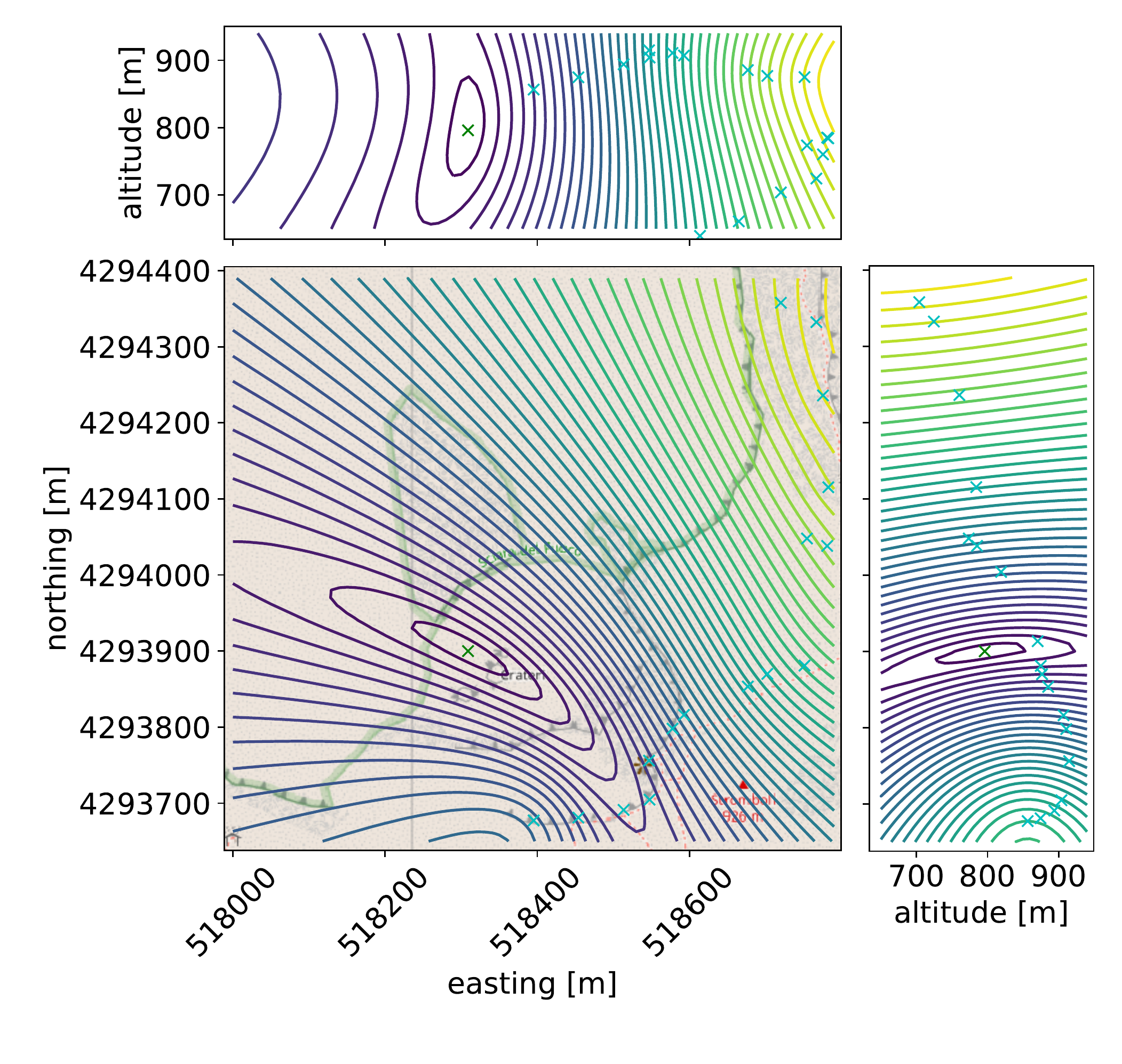}
	\caption{Localisation error for one example eruption. The contour plots show the localisation error surface on the three coordinate planes that intersect with the point of the smallest error, which is the estimate for the source location (the green X in the three plots). In addition, the projections of the module positions onto those planes are plotted as cyan X.}
	\label{fig:rmsMap}
\end{figure*}

\begin{figure*}[ht]
		\centering
		\includegraphics[width=0.9\textwidth]{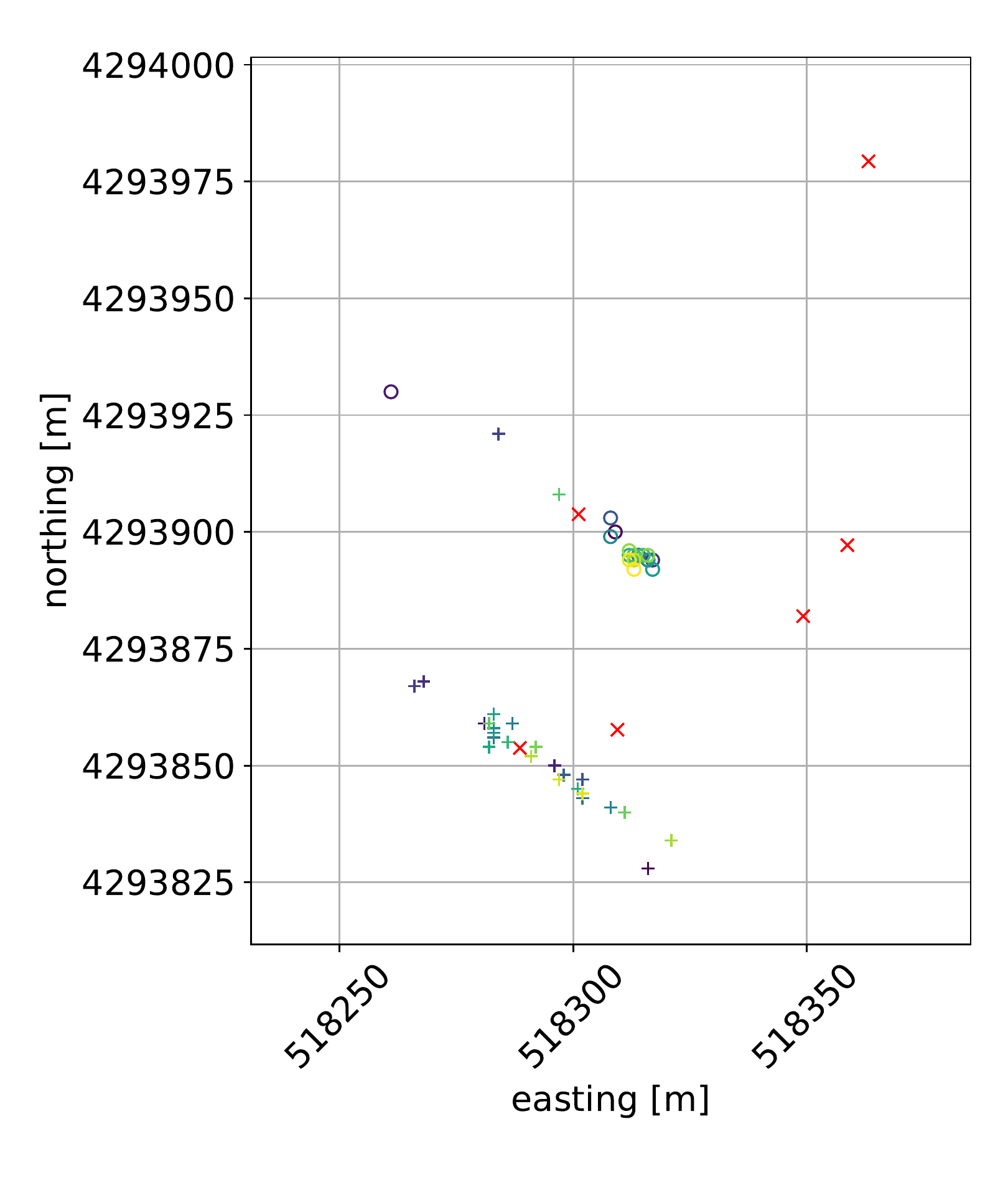}
	\caption{Location of all identified eruptions. Zoomed in view on the crater area, without background map for better visibility. Here, the two categories of the events are marked with "o" for the strong events and with "+" for the weak events. The colour gradient of the markers encodes the chronological order of the events from dark purple (first) to light yellow (last)}
	\label{fig:allEruptions}
\end{figure*}

We analyse all the events recorded on the day of measurement. Figure \ref{fig:rmsMap} shows the exemplary localisation error surface for one event on the three coordinate planes that intersect with point of the smallest error, which is the estimate for the source location. In addition to estimating the location of all identified events, we also classify them into two groups by amplitude (weak and strong events). The threshold for the two classes was chosen to be at 5~\si{\pascal}, as figure \ref{fig:prefilter}, bottom left, shows that some events are well above that threshold and others are well below. The same is true for the rest of the recorded signals that are not shown. Figure \ref{fig:allEruptions} shows the identified location of all the eruptions. Events marked with "o" are above the threshold and events marked with "+" are below. The threshold is applied over all modules, meaning that if any module picked up a sample with a amplitude above 5~\si{\pascal} during an event, it is classified into the strong category. The colour gradient of the markers encode the chronological order of the events from dark purple to light yellow. We compare the results with the location of the vents as identified by image material from a drone, recorded on the same day, marked by the red X. It can be seen that even though the strong simplification of assuming a constant speed of sound of 343~\si{\m\per\s} and no influence by the wind, the locations of the sound sources and thereby the location of the eruptions, matches the positions of the vents quite well. Especially the loud events above the threshold cluster very accurately close to one of the vents, with one exception.\\
However, for the events below the threshold, there is a clear direction of uncertainty visible along the north-west direction. It can be seen in the the localisation error in figure \ref{fig:rmsMap}, as well as in the fact that all identified eruptions lie on a line along the north-west direction in figure \ref{fig:allEruptions}. We conclude that the reason for this is two-fold:\\

\begin{enumerate}
	\item The access to the areas around the summit is strictly restricted due to difficult terrain and the risk of hazard from eruptions. Therefore, the shape in which the modules were placed around the summit, with all modules being east of the crater area, roughly resembles a line along the south west direction. Thus, the uncertainty in the localization along the north-west direction (roughly perpendicular to the line the modules are spread along) is largest, since there is no TDoA information from across the crater area. This can also be seen in figure \ref{fig:rmsMap}, where the area of the minimum localisation error form a valley that spreads along the north west direction. As we assume a constant speed of sound, fluctuations in the actual speed of sound lead to errors that are largest along the north-west direction.
	\item As the events that are spread along the mentioned line are all below the threshold, the SNR of the recorded signals are likely not as high, which in turn lead to errors in determining the TDoA between the modules. Coupled with the fact that the speed of sound is assumed as constant as well as no wind interaction, the errors add up.
\end{enumerate}

The errors in the TDoA that result from a low amplitude of the events lead to a generally wider spread of the localized positions, while the distinct direction of uncertainty that results from the placement of the modules induce that the spread is strongest in that direction, leading to the observed pattern. The offset to the south east that can be observed for the strong events (ignoring the outlier) can be explained by the direction of uncertainty as well, together with a small error in the assumed speed of sound.\\
With this information we conclude that on the day of measurement, only 2 vents were erupting, marked with a red circle in \ref{fig:allEruptions}. Considering the assumptions, these results are considered to be satisfactory accurate, especially as the spread would likely be much smaller if it would have been possible to place modules on the north-west side of the crater area.\\
The results also directly show that on the day of measurement, the erupting vent to north had much stronger eruptions than the one more to the south and that there was no shift over time in activity from one vent to the other.
The acoustic measurements agree with visual observations on the day, suggesting activity of the identified two vents.

\subsection{GPS accuracy} \label{sec:GPSacc}
\begin{figure*}[ht]
	\begin{subfigure}[c]{0.45\textwidth}
		\centering
		\includegraphics[width=1\textwidth]{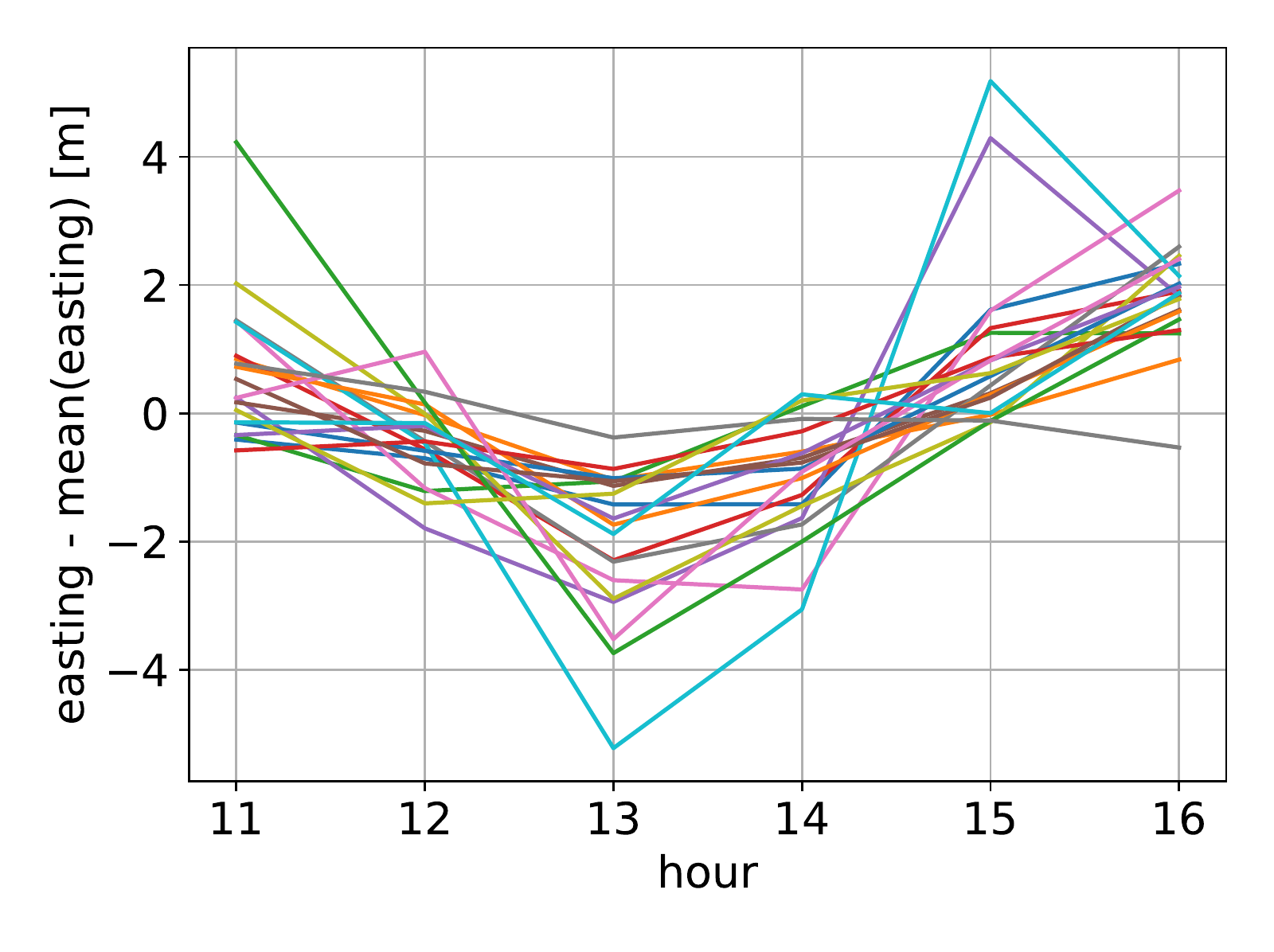}
	\end{subfigure}
	\begin{subfigure}[c]{0.45\textwidth}
		\centering
		\includegraphics[width=1\textwidth]{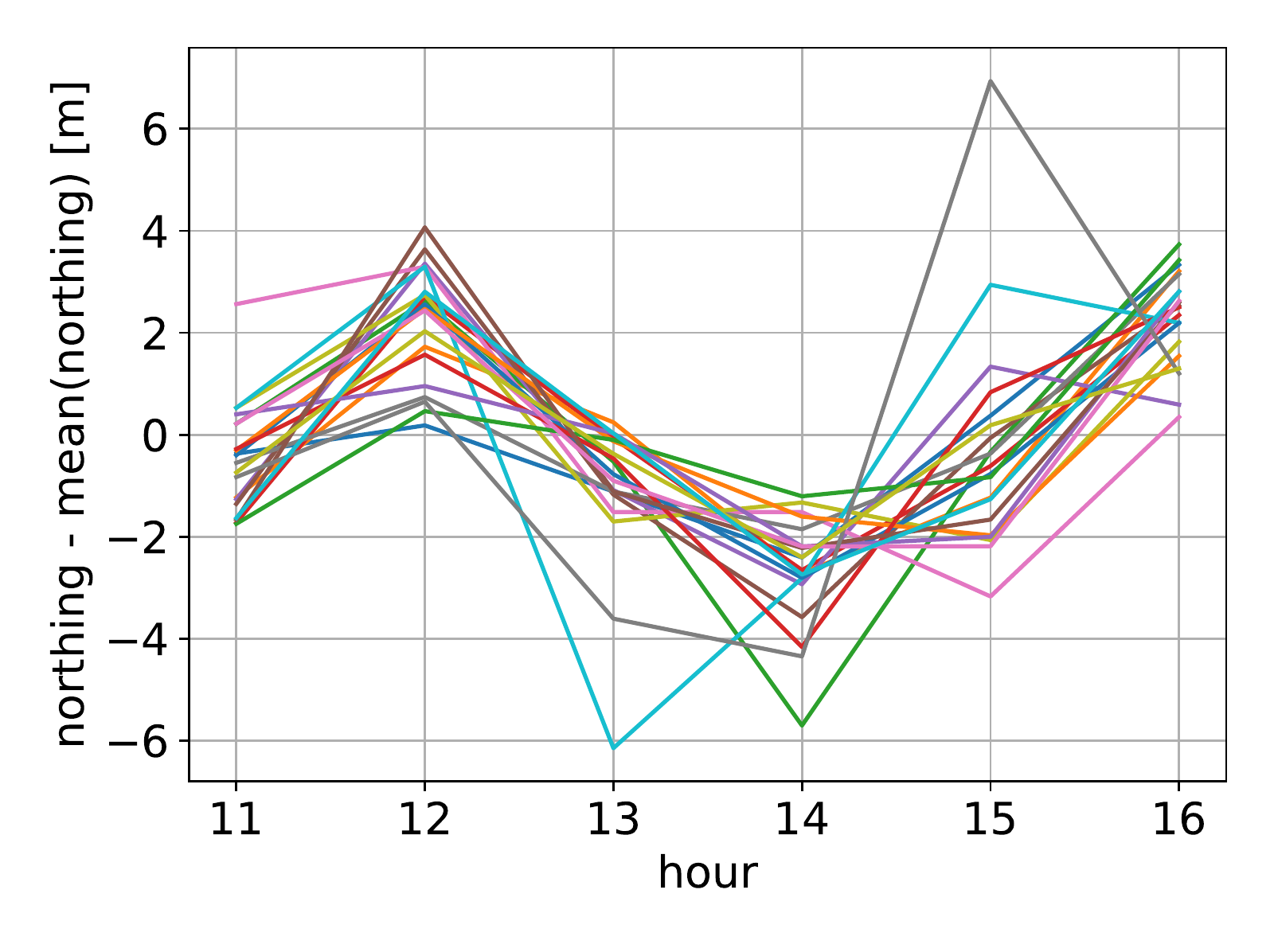}
	\end{subfigure}\\
	\begin{subfigure}[c]{1\textwidth}
		\centering
		\includegraphics[width=0.6\textwidth]{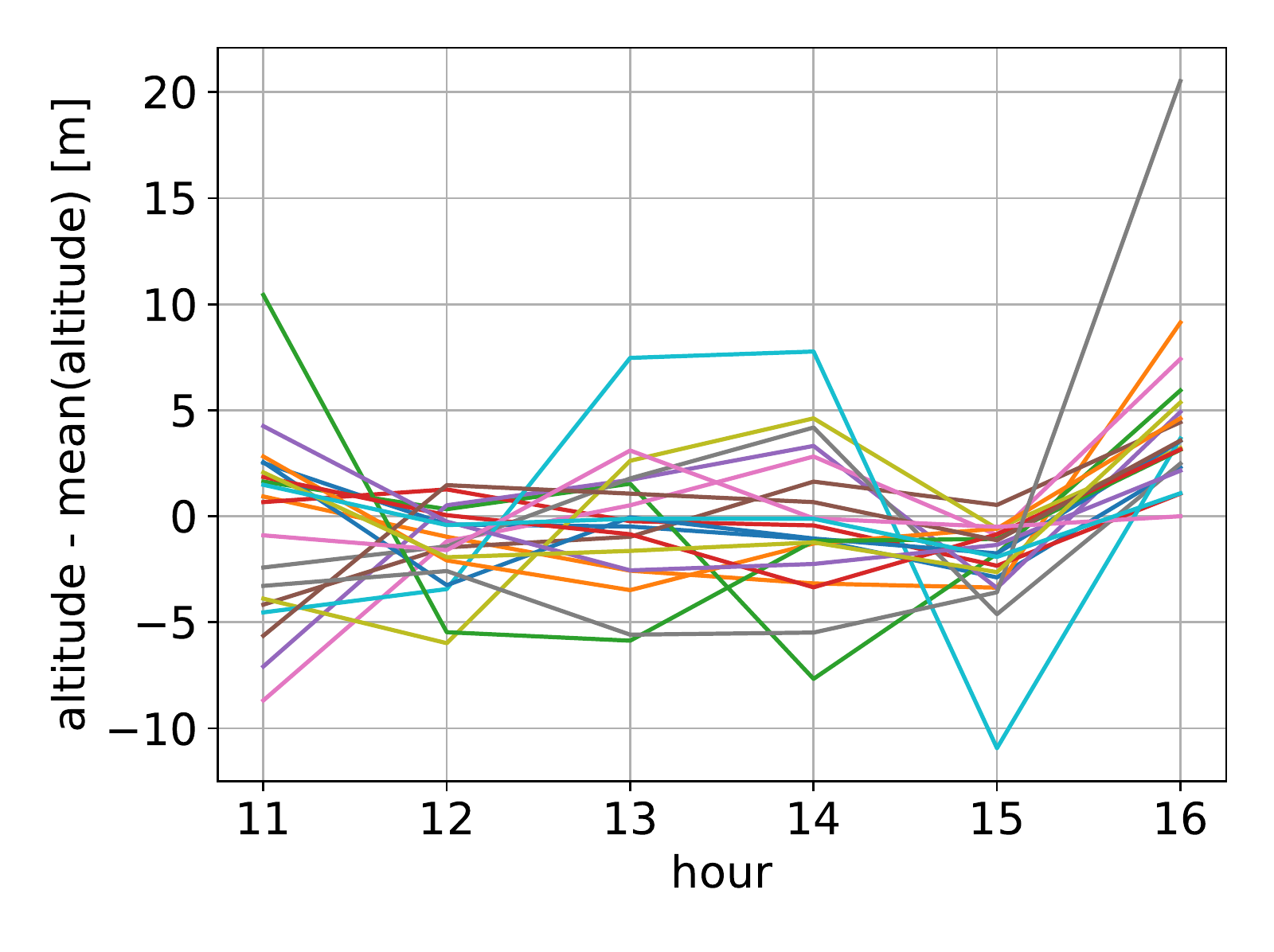}
	\end{subfigure}
	\caption{Fluctuation of GPS position of all modules over time}
	\label{fig:GPSacc}
\end{figure*}

As mentioned in \ref{sec:Hardware}, the best accuracy the Ultimate GPS receivers can achieve is 3~\si{\m}, as given in the corresponding data sheet. As was mentioned as well, the GPS positions were recorded hourly and saved to a text file. In order to improve the accuracy, we use the mean value of all recorded positions of each module as the actual GPS position. Figure \ref{fig:GPSacc} shows the difference of the measured position from the mean over time for all modules. As can be seen for the GPS coordinates, the accuracy of 3~\si{\m} is roughly met, except for a few outliers, whereas for the altitude, the accuracy in the range of around 10~\si{\m} (ignoring the outliers as well). This is to be expected, as the estimation of altitude is usually not as accurate in GPS. What also can be seen is that the fluctuations are very correlated between the modules. From simple visual inspection, the impression arises that the mean values should approximate the true position with a much higher accuracy then 3~\si{\m}.\\
For the application that is shown here, namely localizing the sound sources on a large area of rocky mountain terrain in the open field with wind and temperature- and humidity fluctuation (and therefore fluctuations in the speed of sound) the localization uncertainty well exceeds the uncertainty resulting from the GPS accuracy.

\section{Conclusion}
We designed a prototype for a modular data acquisition system for acoustic field measurements. We manufactured 20 modules and collect useful data on the volcano \textit{Stromboli}, Itay. We demonstrate the system by identifying the source locations of the sound that the eruptions produced. We achieve satisfactory accuracy with a simple method of analysis. The system is very easy to set up and to use. Different settings like the sample rate can be easily controlled through a configuration file that is placed onto the USB storage before the module is started. To deploy a module, it simply needs to be connected to the power supply and placed on the appropriate spot. The cost of one module is around 200€ - 250€, making it much less costly then solutions currently available from industry. The system is especially useful for field campaigns, due to it's light weight and the ease to set up an arbitrary large microphone array, without the need for heavy cables to connect the modules. This becomes even more useful when the distance between microphones is large or the deployment area is not easily accessible. The results of the source identification show one possible application that gives satisfactory results despite simplifications of constant speed of sound in time and space as well as neglecting wind influence. However, we can see that for the purpose of using the modules as an array to localize sound sources with the method described in this work, it would be beneficial for the accuracy to have modules placed around the sound location. Otherwise it may lead to large inaccuracies as a polar pattern in the localization error occurs on the offside of the sound source.\\
It has to be pointed out that the modules are still in a prototype stage. The results still show a good proof of concept to build upon. A multitude of possible improvements come to mind:
\begin{itemize}
	\item The problem that leads to an overvoltage on the microphone supply current needs to be fixed. Adding an additional reference voltage to the PCB design is likely a solution.\\
	\item Adding infrasound capabilities. This could be done in different ways: 
	\begin{itemize}
		\item by simply using a more high-end microphone sensor that can measure infrasound as well. This will rise the cost for each module likely by a lot, since the used microphone sensor now is only around 20€.\\
		\item adding another infrasound sensor that operates in parallel to the acoustic microphone sensor. As there is much room left in the available performance of the rpi3B, this would not be problematic. An option would be to use the sensor that is used in the GEM infrasound data logger \cite{Anderson:2018}. Obviously there is some more work needed here, as the corresponding communication protocol has to be implemented into the custom OS and the PCB design needs to be adapted to incorporate the corresponding parts.\\
	\end{itemize}
	\item To identify wind conditions in the field as well as the speed of sound, acoustic test signals could be send out from different known positions, e.g. from the modules themself\\
	\item Increasing measurement resolution of the acoustic measurements by using an ADC with a higher bit number or a ADC that operates in differential mode\\
	\item Increasing GPS accuracy by exploiting the fact that the fluctuations are mostly correlated between modules. There are also methods that use specific correction signals in differential GPS technology\\
	\item As was shown above, for source localization applications, it is important to place modules around all sides of the source area. As the modules are light weight (around 750 gram, depending on the used power bank, which is by far the heaviest part with around 500 gram in our case) and can be made very robust by according casing, it should be possible without much effort to build a system to place and collect the modules with drones. This would make the modules even more useful in the volcanic setting, where the terrain is often difficult and dangerous.
\end{itemize}


\section{List of Author contributions}
S. Büchholz designed and manufactured the data acquisition system, carried out the measurements and analysed the data. \\
J. Reiss gave support in the layout and design of the electrical circuits.\\
M. Lemke contributed to the sound source localisation method.\\
J.L. Sesterhenn supported in the planning of the experiments and participated in the measurements.\\


\end{document}